\documentclass[12pt]{article}

\usepackage{amsmath,amssymb}

\newcommand{\be}{\begin{equation}}
\newcommand{\ee}{\end{equation}}
\newcommand{\bea}{\begin{eqnarray}}
\newcommand{\eea}{\end{eqnarray}}
\newcommand{\nn}{\nonumber \\}
\newcommand{\p}[1]{(\ref{#1})}
\newcommand{\lb}{\label}

\newcommand{\und}{\underline}

\begin{document}
\begin{titlepage}
\begin{flushright}
\end{flushright}
\vskip 0.6truecm

\begin{center}
{\Large\bf Biharmonic Superspace for ${\cal N}{=}4$
Mechanics}
\vspace{1.5cm}

{\large\bf E. Ivanov$\,{}^a$, J. Niederle$\,{}^b$}\\
\vspace{1cm}

{\it a)Bogoliubov  Laboratory of Theoretical Physics, JINR,}\\
{\it 141980 Dubna, Moscow region, Russia} \\
{\tt eivanov@theor.jinr.ru}\vspace{0.3cm}\\
{\it b)Institute of Physics of the Academy of Science of the Czech Republic,}\\
{\it Prague 8, CZ 182 21, Czech Republic}\\
{\tt niederle@fzu.cz} \vspace{0.3cm}
\end{center}
\vspace{0.2cm}
\vskip 0.6truecm  \nopagebreak

\begin{abstract}
\noindent We develop a new superfield approach to ${\cal N}{=}4$
supersymmetric mechanics based on the concept of biharmonic
superspace (bi-HSS). It is an extension of the ${\cal N}{=}4,
d{=}1$ superspace by two sets of harmonic variables associated
with the two SU(2) factors of the R-symmetry group SO(4) of the
${\cal N}{=}4, d{=}1$ super Poincar\'e algebra. There are three
analytic subspaces in it: two of the Grassmann dimension 2 and one
of the dimension 3. They are closed under the infinite-dimensional
``large'' ${\cal N}{=}4$ superconformal group, as well as under
the finite-dimensional superconformal group $D(2,1;\alpha)$. The
main advantage of the bi-HSS approach is that it gives an
opportunity to treat ${\cal N}{=}4$ supermultiplets with finite
numbers of off-shell components on equal footing with their
``mirror'' counterparts. We show how such multiplets and their
superconformal properties are described in this approach. We also
define nonpropagating gauge multiplets which can be used to gauge
various isometries of the bi-HSS actions. We present an example of
nontrivial ${\cal N}{=}4$ mechanics model with a seven-dimensional
target manifold obtained by gauging an $\rm{U}(1)$ isometry in a
sum of the free actions of the multiplet ${\bf (4,4,0)}$ and its
mirror counterpart.
\end{abstract}
\vspace{0.7cm}

\noindent PACS: 11.30.Pb, 11.15.-q, 11.10.Kk, 03.65.-w\\
\noindent Keywords: Supersymmetry, gauging, isometry, superfield
\newpage

\end{titlepage}

\section{Introduction}

Supersymmetric models in one dimension (i.e. models of supersymmetric
mechanics) present an interesting subject of study. The main reason for the
interest in such models is founded on the point that after quantization they yield the
supersymmetric quantum mechanics which can be efficiently used to better
understand  the characteristic features of higher-dimensional
supersymmetric field theories, e.g. the various mechanisms of
spontaneous breaking of supersymmetry \cite{Wit}. These models,
especially their superconformally invariant versions \cite{AP} - \cite{BMSV}, also bear
intimate relationships with superparticles, black holes, and
gauge/string correspondence. Leaving aside the connection with
higher-dimensional theories (see, e.g., \cite{ASm} for a review)
and superbranes, the supersymmetric
mechanics possesses a number of surprising properties and
applications {\it per se}. For instance, it is an exciting task to
construct supersymmetric extensions of some important intrinsically
one-dimensional models, such as the integrable Calogero and
Calogero-Moser systems, various generalizations of the renowned
Landau problem, the outstanding quantum Hall effect,
etc (see, e.g., \cite{Freed} - \cite{Poly} and references therein).

Moreover, one-dimensional extended supersymmetry reveal many specific unusual
features which are not inherent in its higher-dimensional
counterparts. For instance, some on-shell multiplets of the latter
become off-shell in $d{=}1$, most of the standard linear constraints
on $d{=}1$ superfields  (e.g., the chirality constraints) have their
nonlinear counterparts giving rise to the intrinsic nonlinearity of
the relevant off-shell supersymmetry transformation laws, etc. The
specific $d{=}1$ phenomenon is the so-called automorphic duality
which relates the off-shell supermultiplets having the same number
of fermionic fields but revealing different divisions of the bosonic fields
into the physical and auxiliary subsets \cite{GR0} - \cite{T}. As shown, for instance, in
\cite{root, DI,DI1,DI2}, all multiplets of ${\cal N}{=}4, \; d{=}1$ supersymmetry
with four physical fermionic fields can be generated from the ``root''
multiplet $(\bf{4, 4, 0})$ via this type of duality.\footnote{From now on, the notation
$(\bf{n_1, n_2, n_3})$ means an off-shell multiplet with ${\bf n_1}$ physical bosonic fields,
${\bf n_2}$ physical fermionic fields and ${\bf n_3} = {\bf n_2 - n_1}$
auxiliary bosonic fields.}

In order to better understand  these and some other remarkable properties of
extended $d=1$ supersymmetries and related supersymmetric mechanics models, it is
useful to have adequate superfield techniques making manifest
as many involved (super)symmetries as possible. It was argued
in \cite{IL,DI,DI1,DI2,DI3} that for ${\cal N}{=}4, \; d{=}1$ supersymmetry such
an underlying superfield approach is the one based on ${\cal N}{=}4, \; d{=}1$
harmonic superspace \cite{DS,DI0}. In particular, the automorphic duality was shown to
be associated with gaugings of various symmetries realized on a specific harmonic superfield
describing the root multiplet $(\bf{4, 4, 0})$.

The harmonic superspace used in \cite{IL} - \cite{DI3} is a direct $d{=}1$ counterpart of the ${\cal N}{=}2, d{=}4$
harmonic superspace \cite{HSS,book}. It involves the harmonic variables associated
with one of the two SU(2) automorphism (or $R$-symmetry) groups of the ${\cal N}{=}4, \; d{=}1$
superalgebra. In this approach another SU(2) is not manifest. On the other hand,
all ${\cal N}{=}4, \; d{=}1$ supermultiplets exist in two forms which differ in the assignment
of the component fields with respect to these two different automorphism SU(2)
symmetries. Since the ${\cal N}{=}4$ multiplets and their ``mirror'' versions are different up to an interchange
of the two SU(2) groups, this difference is not essential when dealing with only
one multiplet from such a pair. When both sorts of the ${\cal N}{=}4, \; d{=}1$ multiplets
are involved (as is the case, e.g., in ${\cal N}{=}8$ mechanics models
formulated in terms of ${\cal N}{=}4$ multiplets \cite{BIKL,ILS}), this difference becomes
essential. It is desirable to have {\it both} automorphism SU(2) symmetries realized in a {\it manifest} way,
in order to efficiently control their breaking patterns, etc.

In this paper, we present basic elements of the appropriate superfield approach. It is based on the notion of
biharmonic superspace which involves the harmonic variables associated with both SU(2)
groups. This type of harmonic superspace was previously used in two dimensions to describe various
types of twisted ${\cal N}{=}(4, 4)$ multiplets and their interactions \cite{IS,BI}, as well as in
${\cal N}{=}8, d{=}1$ supersymmetry \cite{BIS}.

The basic definitions are introduced in Sect. 2. In particular, we show that the ${\cal N}{=}4$
bi-HSS involves three different analytic subspaces: two subspaces of Grassmann dimension 2 (i.e.
with the two Grassmann coordinates) and one of Grassmann dimension 3. In Sect. 3 we show that all
three analytic subspaces are closed under the appropriate realizations of the infinite-dimensional ``large''
${\cal N}{=}4$ superconformal group. The corresponding coordinate transformations, as well as
those of the finite-dimensional ${\cal N}{=}4$ superconformal group $D(2,1;\alpha)$, are specified. In Sect. 4
we present the bi-HSS description of the basic off-shell ${\cal N}{=}4$ supermultiplets with
four physical fermions, and discuss peculiarities of the relevant realizations of the ${\cal N}{=}4$ superconformal
groups. We also give an example of the new ${\cal N}{=}4$ multiplet with the off-shell content $(8 +8)$
described by a superfield living on the three-theta analytic superspace. The relevant invariant actions yield
 a Wess-Zumino (WZ)-type term of the first order in the time derivative for physical bosons, as opposed
to the standard second-order
kinetic term. In Sect. 5 we discuss nonpropagating gauge multiplets in the ${\cal N}{=}4$ bi-HSS that
allow one to gauge isometries of the matter ${\cal N}{=}4, d{=}1$ actions while preserving the relevant harmonic
analyticities. An example of such a gauged model with a seven-dimensional bosonic target manifold is presented.
Its component action contains, besides the sigma-model-type term, a scalar potential and a WZ term.

\setcounter{equation}{0}

\setcounter{equation}{0}
\section{The ${\cal N}{=}4\,, \; d{=}1$ biharmonic superspace: basics}
We begin with the ordinary ${\cal N}{=}4, d{=}1$ superspace in a notation with
the both SU(2) automorphism groups being manifest. It is defined as the coordinate set
\be
z = (t,\theta^{ia})\,,\lb{N40}
\ee
in which the ${\cal N}{=}4, d{=}1$ supersymmetry is realized by means of the transformations
\be
\delta t = -i\varepsilon^{ia}\theta_{ia}\,, \quad \delta \theta^{ia} = \varepsilon^{ia}. \lb{N4}
\ee
The Grassmann coordinates $\theta^{ia}$ (as well as the supertranslations parameters $\varepsilon^{ia}$)
form a real quartet of the full automorphism
group SO(4) $\sim$ SU(2)$_L\times $SU(2)$_R$,
$\overline{(\theta^{ia})} = \theta_{ia} = \epsilon_{ik}\epsilon_{ab}\theta^{kb}\,$. The indices $i$ and $a$ are
doublet indices of the left and the right SU(2) automorphism groups, respectively.
The corresponding covariant spinor derivatives are defined as
\begin{eqnarray}
&& D_{ia}=\frac{\partial}{\partial \theta^{ia}}+i\theta_{ia}\partial_{t}\,,\quad
\bar D^{ia}= -\frac{\partial}{\partial \theta_{ia}}- i\theta^{ia}\partial_{t} = -\epsilon^{ik}\epsilon^{ab} D_{kb}\,,\nn
&& \{D_{ia}, D_{kb}\}=2i\epsilon_{ik}\epsilon_{ab}\partial_{t}\,.
\end{eqnarray}

In the {\it central} basis, the ${\cal N}{=}4,\; d{=}1$ biharmonic superspace (bi-HSS)
is defined as the following extension of \p{N40}
\be
(z, u, v) = (t,\;\theta^{ia}\,,\; u^{\pm 1}_i\,, \; v^{\pm 1}_b)\,.
\ee
Here, $u^{\pm 1}_i \in {\rm SU}(2)_L/{\rm U}(1)_L\,$ and $v^{\pm 1}_a \in {\rm SU}(2)_R/{\rm U}(1)_R\,$ are two independent sets
of SU(2) harmonic variables. The harmonics $u^{\pm 1}_i$ satisfy the standard relations \cite{HSS,book}
\be
u^{-1}_i = \overline{(u^{1\,i})}\,, \quad u^{1\,i}u^{-1}_{i}=1\; \Leftrightarrow \;
u^1_iu^{- 1}_k - u^{1}_k u^{-1}_i = \epsilon_{ik}\,. \lb{CompL}
\ee
The same relations are valid for harmonics $v^{\pm 1}_a$, with the change $i,k \rightarrow a, b$. Though we denote the harmonic
charges of $u$ and $v$ by the same indices, these charges are completely independent. So the harmonic part of the biharmonic superspace
is the coset $\frac{{\rm SU}(2)_L}{{\rm U}(1)_L}\otimes \frac{{\rm SU}(2)_R}{{\rm U}(1)_R}$. As usual, the fact that
all biharmonic superfields, i.e., $\Phi^{(q, p)}(z,u,v)$,
are defined just on this coset is expressed as requirement that both harmonic $\rm{U}(1)$ charges are strictly preserved
in all superfield actions. For superfields $\Phi^{(q,p)}(z,u,v)\,$, we assume a double harmonic expansion over
the harmonic monomials which are constructed from $u^{\pm 1}_i$ and $v^{\pm 1}_a\,$, respectively,
and which have $\rm{U}(1)$ charges $q$ and $p\,$.

A specific feature of the ${\cal N}{=}4, \;d{=}1$ bi-HSS is the existence of two types of {\it analytic} bases with
the {\it analytic} subspaces having half of the Grassmann variables, as compared to the full Grassmann dimension 4 of
the bi-HSS.  These two analytic bases are spanned by the following coordinate sets
\begin{eqnarray}
&&(z_+, u, v) = \left( t_{+}=t+ i(\theta^{1,1}\theta^{-1, -1} +\theta^{-1, 1}\theta^{1, -1})\,,\;
\theta^{1,1}, \theta^{1,-1}, \theta^{-1,1}, \theta^{-1,-1}, u^{\pm 1}_{i}\,, v^{\pm 1}_{a}\right), \lb{A+}\\
&&(z_-, u, v) = \left( t_{-}=t+ i(\theta^{1,1}\theta^{-1, -1} -\theta^{-1, 1}\theta^{1, -1})\,,\;
\theta^{1,1}, \theta^{1,-1}, \theta^{-1,1}, \theta^{-1,-1}, u^{\pm 1}_{i}\,, v^{\pm 1}_{a}\right), \lb{A-}
\end{eqnarray}
where
\be
\theta^{m, n} := \theta^{ia}u^m_i v^n_a\,, \quad m, n = \pm 1\,. \lb{DefPr}
\ee

Defining harmonic projections of the spinor derivatives as
\bea
&& D^{m, n} = D^{ia}u^m_i v^n_a\,, \lb{DefD1} \\
&& (D^{1,1})^2 = (D^{-1,-1})^2 = (D^{1,-1})^2 = (D^{-1,1})^2 = \{D^{\pm 1,1}, D^{\pm 1,-1}\} = \{D^{1,\pm 1}, D^{-1,\pm 1}\} = 0\,, \nn
&& \{D^{1,1}, D^{-1,-1}\} = -\{D^{1,-1}, D^{-1,1}\} = 2i\partial_t\,, \lb{Dcomm}
\eea
it is easy to show that, in the above bases, they have the form
\bea
&& D^{1,1} = \frac{\partial}{\partial \theta^{-1,-1}}\,, \quad D^{1,-1} = -\frac{\partial}{\partial \theta^{-1,1}}\,, \nn
&& D^{-1,1} = -\frac{\partial}{\partial \theta^{1,-1}} + 2i\theta^{-1,1}\partial_{t_+}\,,
\quad D^{-1,-1} = \frac{\partial}{\partial \theta^{1,1}} + 2i\theta^{-1,-1}\partial_{t_+}\,, \lb{Cov+}
\eea
and
\bea
&& D^{1,1} = \frac{\partial}{\partial \theta^{-1,-1}}\,, \quad D^{-1,1} = -\frac{\partial}{\partial \theta^{1,-1}}\,, \nn
&& D^{1,-1} = -\frac{\partial}{\partial \theta^{-1,1}} + 2i\theta^{1,-1}\partial_{t_-}\,,
\quad D^{-1,-1} = \frac{\partial}{\partial \theta^{1,1}} + 2i\theta^{-1,-1}\partial_{t_-}\,. \lb{Cov++}
\eea
The fact that two different pairs of covariant spinor derivatives are reduced to the partial derivatives
in these bases implies the existence of two
analytic subspaces which are closed under the full ${\cal N}{=}4$ supersymmetry.
Hence there are two sorts of analytic superfields defined as
unconstrained functions on these analytic superspaces:
\bea
&&(\zeta_+, u, v) = \left( t_{+}, \theta^{1,1}, \theta^{1,-1}, u^{\pm 1}_{i}\,, v^{\pm 1}_{a}\right), \lb{Asubs+} \\
&& D^{1,1}\Phi_{(+)} = D^{1,-1}\Phi_{(+)} = 0\quad \Rightarrow \quad \Phi_{(+)} = \varphi_{(+)}(\zeta_+, u, v)\,, \lb{Anal+}
\eea
and
\bea
&&(\zeta_-, u, v) = \left( t_{-}, \theta^{1,1}, \theta^{-1,1}, u^{\pm 1}_{i}\,, v^{\pm 1}_{a}\right), \lb{Asubs-} \\
&& D^{1,1}\Phi_{(-)} = D^{-1,1}\Phi_{(-)} = 0\quad \Rightarrow \quad \Phi_{(-)} = \varphi_{(-)}(\zeta_-, u, v)\,. \lb{Anal-}
\eea

An important role in the harmonic superspace approach is played by harmonic derivatives. The harmonic derivatives with respect
to harmonics $u^{\pm 1}_i$ and $v^{\pm 1}_a$ in the central basis are defined as
\be
\partial^{\pm 2, 0} = u^{\pm 1}_i\frac{\partial}{\partial u^{\mp 1}_i}\,, \quad
\partial^{0, \pm 2} = v^{\pm 1}_a\frac{\partial}{\partial v^{\mp 1}_a}\,.
\ee
These sets form two mutually commuting SU(2) algebras
\bea
&&[\partial^{2, 0},\partial^{-2, 0}] = \partial_u^0\,, \; [\partial_u^0, \partial^{\pm 2, 0}]  = \pm 2 \partial^{\pm 2, 0}\,, \nn
&&\;[\partial^{0,2},\partial^{0,-2}] = \partial_v^0\,,\;
\;[\partial_v^0, \partial^{0, \pm 2}]  = \pm 2 \partial^{0, \pm 2}\,,\nn
&& [\partial^{\pm 2, 0},\partial^{0, \pm 2}] =[\partial^{\pm 2, 0},\partial^{0, \mp 2}] =[\partial^{\mp 2, 0},\partial^{0, \pm 2}] = 0\,.
\lb{HarmComm}
\eea
Here,
\be
\partial_u^0 = u^{1}_i\frac{\partial}{\partial u^{1}_i} - u^{-1}_i\frac{\partial}{\partial u^{-1}_i}\, \;\mbox{and}\; \,
\partial_v^0 = v^{1}_a\frac{\partial}{\partial v^{1}_a} - v^{-1}_a\frac{\partial}{\partial v^{-1}_a}
\ee
are operators corresponding to the two independent external harmonic $\rm{U}(1)$ charges. In the analytic bases \p{A+} and \p{A-}
the harmonic derivatives acquire additional terms. For instance, in basis \p{A+}:
\bea
&&D^{\pm 2, 0} = \partial^{\pm 2, 0} \pm 2i \theta^{\pm 1, \pm 1}\theta^{\pm 1, \mp 1}\partial_{t_+}
+ \theta^{\pm 1, \pm 1}\frac{\partial }{\partial \theta^{\mp 1, \pm 1}}
+ \theta^{\pm 1, \mp 1}\frac{\partial }{\partial \theta^{\mp 1, \mp 1}}\,, \nn
&&D^{0, \pm 2}= \partial^{0, \pm 2}+ \theta^{\pm 1, \pm 1}\frac{\partial }{\partial \theta^{\pm 1, \mp 1}}
+ \theta^{\mp 1, \pm 1}\frac{\partial }{\partial \theta^{\mp 1, \mp 1}}\,, \nn
&&D^{0}_u =\partial^{0}_u + \left(\theta^{1, 1}\frac{\partial }{\partial \theta^{1, 1}}
+ \theta^{1, -1}\frac{\partial }{\partial \theta^{1, -1}} - \theta^{-1, 1}\frac{\partial }{\partial \theta^{-1, 1}} -
\theta^{-1, -1}\frac{\partial}{\partial \theta^{-1, -1}}\right), \nn
&&D^{0}_v =\partial^{0}_v + \left(\theta^{1, 1}\frac{\partial }{\partial \theta^{1, 1}}
+ \theta^{-1, 1}\frac{\partial }{\partial \theta^{-1, 1}} - \theta^{1, -1}\frac{\partial }{\partial \theta^{1, -1}} -
\theta^{-1, -1}\frac{\partial }{\partial \theta^{-1, -1}}\right). \lb{DHanalB}
\eea
Their commutation relations are given again by the same formulas \p{HarmComm} because they are basis-independent.

For what follows, it is important to know the commutators of the harmonic derivatives with the spinor ones.
Independently
of the basis, these commutation relations are
\bea
&& [D^{2,0}, D^{1,1}] = [D^{2,0}, D^{1,-1}] = 0\,, \; [D^{2,0}, D^{-1,1}] = D^{1,1}\,, \;[D^{2,0}, D^{-1,-1}] = D^{1,-1}\,, \lb{I} \\
&& [D^{-2,0}, D^{-1,1}] = [D^{-2,0}, D^{-1,-1}] = 0\,, \nn
&& [D^{-2,0}, D^{1,1}] = D^{-1,1}\,, \;[D^{-2,0}, D^{1,-1}] = D^{-1,-1}\,, \lb{II} \\
&& [D^{0,2}, D^{1,1}] = [D^{0,2}, D^{-1,1}] = 0\,, \; [D^{0,2}, D^{1,-1}] = D^{1,1}\,, \;[D^{0,2}, D^{-1,-1}] = D^{-1,1}\,, \lb{III} \\
&& [D^{0,-2}, D^{1,-1}] = [D^{0,-2}, D^{-1,-1}] = 0\,, \nn
&& [D^{0,-2}, D^{1,1}] = D^{1,-1}\,, \;[D^{0,-2}, D^{-1,1}] = D^{-1,-1}\,. \lb{IV}
\eea
The commutation relations involving the harmonic-charge operators $D^0_u$ and $D^0_v$ are evident:
$$
[D^0_u, D^{m,n}] = m\, D^{m,n}\,, \;[D^0_v, D^{m,n}] = n\, D^{m,n}\,.
$$
{}From all these relations we observe another unusual property of the ${\cal N}{=}4, d{=}1$ bi-HSS, namely, that each
of the sets of analyticity conditions \p{Anal+} and \p{Anal-} is preserved by three harmonic derivatives. These are
$D^{2,0}, D^{0,2}, D^{0,-2}$ and $D^{2,0}, D^{0,2}, D^{-2,0}$, respectively. This is due to the fact that
the spinor derivatives entering \p{Anal+} and \p{Anal-},
together with these harmonic derivatives, form closed subalgebras (the so-called Cauchi-Riemann (${\bf CR}$) structures):
\bea
&&{\bf CR}_{(+)} = \left( D^{1,1}, D^{1,-1}, D^{2,0}, D^{0,2}, D^{0, -2}\right), \lb{CR1} \\
&&{\bf CR}_{(-)} = \left( D^{1,1}, D^{-1,1}, D^{2,0}, D^{0,2}, D^{-2,0}\right). \lb{CR2}
\eea
The homogeneously acting $\rm{U}(1)$ charge operators $D^0_u$ and $D^0_v\,$ should be added to these sets.

To finish the discussion concerning the ${\cal N}{=}4$ biharmonic analyticities, we note
that the two-theta analytic subspaces \p{Asubs+} and \p{Asubs-} can be embedded
into an analytic subspace with three theta-coordinates. In basis \p{A+}, it is given by the following set of coordinates
\be
(\zeta, u, v) = \left(t_+\,, \theta^{1,1}\,, \theta^{1,-1}\,, \theta^{-1, 1}\,, u^{\pm 1}_{i}\,, v^{\pm 1}_{a}\right),\lb{Asubs3}
\ee
which corresponds to imposing the relaxed Grassmann analyticity condition on the bi-biharmonic superfields:
\be
D^{1,1}{\Phi}_{(3)} = 0 \; \Rightarrow \; \Phi_{(3)} = \varphi_{(3)}(\zeta, u, v)\,.\lb{Weak}
\ee
This analyticity is preserved only by two harmonic derivatives commuting with $D^{1,1}$, i.e. by $D^{2,0}$ and $D^{0,2}\,$.
So, the corresponding ${\bf CR}$ structure is
\be
{\bf CR}_{(3)} = \left( D^{1,1}, D^{2,0}, D^{0,2} \right). \lb{CR3}
\ee

An important notion of the harmonic approach is the so-called ``tilde conjugation,'' which is a product
of the ordinary complex conjugation and the Weyl reflection (antipodal map) on the harmonic sphere $S^2$ \cite{HSS,book}.
In our case, the basic rules of the tilde conjugation are
\be
\widetilde{\theta^{p,q}} = \theta^{p,q}\,, \; \widetilde{u^{\pm 1}_i} = u^{\pm 1 i}\,, \;\widetilde{v^{\pm 1}_a}  = v^{\pm 1 a}\,, \;
\widetilde{t_\pm} = t_{\pm} \; \mbox{and} \; \widetilde{t} = t\,. \lb{tilda}
\ee
Being applied twice, this involution yields $-1$ on the harmonic variables, but $+1$ on the harmonic projections of the
Grassmann coordinates due to the presence of two sets of harmonics in their definition \p{DefPr}
(as distinct from the case of harmonic superspaces with one set of
harmonics \cite{HSS,IL}). The analytic subspaces \p{Asubs+}, \p{Asubs-} and \p{Asubs3} are closed under this conjugation (as opposed,
e.g., to the chiral ${\cal N}{=}4$ superspaces which are not closed under the complex conjugation)
and thus one can choose the corresponding analytic superfields to be real with respect to it.

For further purposes, we shall also need the following important statement concerning functions expandable in the double harmonic series
in $u^{\pm 1}_i$ and $v^{\pm 1}_a\,$.\\

\noindent {\it Proposition.} -- For all biharmonic functions $B^{(q, p)}$ with $p < 0$ and $C^{(q,p)}$ with $q < 0 $ the following statements hold:
\bea
&& D^{0,2}B^{(q, p)} = 0 \;\;\Rightarrow \;\; B^{(q, p)} = 0 \,, \lb{Lemma1} \\
&& D^{2,0}C^{(q, p)} = 0 \;\;\Rightarrow \;\; C^{(q, p)} = 0 \;\,. \lb{Lemma2}
\eea
They can be proved by expanding $B^{(q, p)}$ and $C^{(q,p)}$ in a biharmonic series, as in the case of the
standard harmonic superspace \cite{HSS}. Similar statements are also true with harmonic
derivatives $D^{-2,0}$ and $D^{0,-2}$, namely that
$$
D^{-2,0}\hat{C}^{(q, p)} = 0 \;\Rightarrow \; \hat{C}^{(q,p)} = 0\,,\;\; \mbox{iff}\;\; q>0 \quad \mbox{and} \quad
D^{0,-2}\hat{B}^{(q, p)} = 0 \;\Rightarrow \; \hat{B}^{(q,p)} = 0\,,\;\;  \mbox{iff} \;\; p>0\,.
$$

Finally, let us define integration measures on the full ${\cal N}{=}4, d{=}1$ bi-HSS and on its analytic subspaces:
\bea
&&\underline{\mbox{Full bi-HSS}}: \quad \int \mu := \int dt\, du\, dv\, (D^{-1,-1}D^{-1,1}D^{1,1}D^{1,-1})\,, \lb{FullMea} \\
&&\underline{\mbox{Analytic bi-HSS 1}}:\quad \int \mu_{A+}^{(-2,0)} := \int dt_+\, du\, dv\, (D^{-1,-1}D^{-1,1})\,, \lb{AMea+} \\
&&\underline{\mbox{Analytic bi-HSS 2}}:\quad \int \mu_{A-}^{(0,-2)} := \int dt_-\, du\, dv\, (D^{-1,-1}D^{1,-1})\,, \lb{AMea-} \\
&&\underline{\mbox{Analytic bi-HSS 3}}:\quad \int \mu_{A3}^{(-1,-1)} := \int dt_+\, du\, dv\, (D^{-1,-1}D^{1,-1}D^{-1,1})\,. \lb{AMea3}
\eea
They are normalized in such a way that
\bea
&& \int \mu \;(\theta^{-1,-1}\theta^{-1,1}\theta^{1,1}\theta^{1,-1}) \times... = \int dt du dv \times...\,, \nn
&& \int \mu_{A+}^{(-2,0)} \;(\theta^{1,1}\theta^{1,-1}) \times... = \int dt_+ du dv \times...\,, \nn
&& \int \mu_{A-}^{(0,-2)} \;(\theta^{1,1}\theta^{-1,1}) \times... = \int dt_-du dv \times...\,, \nn
&& \int \mu_{A3}^{(-1,-1)} \;(\theta^{1,1}\theta^{1,-1}\theta^{-1,1}) \times... = \int dt_+du dv \times... = \int dt_-du dv \times... \nonumber
\eea
(up to a total time derivative).
\setcounter{equation}{0}

\section{${\cal N}{=}4$ superconformal groups}

By analogy with the ${\cal N}{=}(2,2), d{=}2$ bi-HSS \cite{IS,BI}, in the ${\cal N}{=}4, d{=}1$ biharmonic superspace one can define
various superdiffeomorphism groups preserving a given type of the harmonic Grassmann analyticity. The resulting gauge theories will correspond
to some versions  of nonpropagating ${\cal N}{=}4, d{=}1$ supergravities which can be used to construct various models
of superparticles in the ${\cal N}{=}4$ bi-HSS (for instance, along the line of refs. \cite{BI,BIS}). Leaving this interesting, but difficult,
problem for the future, in this section we are interested  in those subgroups of the
general diffeomorphism group in a biharmonic ${\cal N}{=}4$
superspace which ({\bf i}) preserve the biharmonic analyticity and ({\bf ii})
do not affect the flat form of the analyticity-preserving harmonic
derivatives. By analogy with the previously known examples, one can
anticipate that these subgroups contain the appropriate ${\cal N}{=}4$
superconformal transformations. This is indeed the case, and
we shall present below the precise form of these transformations.

\subsection{Infinite-dimensional ${\cal N}{=}4$ superconformal groups}

In our search for the analyticity-preserving realizations of ${\cal N}{=}4$ superconformal groups
we shall consider both the three-theta analytic
subspace \p{Asubs3} and one of the two-theta analytic subspaces \p{Asubs+} (realizations in the ``mirror'' subspace \p{Asubs-}
are obtained via the substitution $t_+ \rightarrow t_-$, $u^\pm_i \leftrightarrow v^\pm_a$ and via the appropriate
substitutions for the odd coordinates).
For convenience, we shall always use the analytic basis \p{A+}.

We start with the following most general coordinate transformations preserving the
three-theta analytic subspace \p{Asubs3}:
\bea && \delta t_+ =
\Lambda(\zeta, u, v), \, \delta \theta^{1,1} =
\Lambda^{1,1}(\zeta, u, v), \,
\delta \theta^{1,-1} = \Lambda^{1,-1}(\zeta, u, v), \,\delta \theta^{-1,1} = \Lambda^{-1,1}(\zeta, u, v), \lb{TranCooan} \\
&&\delta u^{1}_i = \Lambda^{2,0}(\zeta, u, v)u^{-1}_i\,, \;\delta u^{-1}_i = 0\,,
\quad\delta v^{1}_a = \Lambda^{0,2}(\zeta, u, v)v^{-1}_a\,, \;\delta v^{-1}_a = 0\,,\lb{TranHarm} \\
&&\delta \theta^{-1,-1} = \Lambda^{-1,-1}(\zeta, \theta^{-1,-1}, u, v)\,. \lb{TranNonan}
\eea
The ``triangular'' form of transformations of harmonic variables \p{TranHarm} (respecting only
the generalized $\widetilde{\;\;\;\;}$ conjugation) has been chosen by analogy with the previously known
examples \cite{HSS}.\footnote{The necessity of just this form of the transformations
of the harmonic variables was further explained  in \cite{Howe}, in the equivalent language of the {\it active}
transformations on superfields and with using
the specific parametrization for the harmonic variables.} The infinitesimal
transformation of the nonanalytic coordinate $\theta^{-1,-1}$ can bear dependence on all coordinates
of the bi-HSS.

Let us postulate now that under these transformations the analyticity-preserving harmonic derivatives
$D^{2,0}$ and $D^{0,2}$ transform as follows:
\be
(\mbox{a})\;\; \delta D^{2,0} = -\Lambda^{2,0}D^0_u\,, \quad (\mbox{b})\;\;\delta D^{0,2} = -\Lambda^{0,2}D^0_v\,. \lb{DTranCo}
\ee
Since $\delta D^0_u = \delta D^0_v =0$, the transformations respecting \p{DTranCo} form a subgroup of the superdiffeomorphism
group \p{TranCooan} - \p{TranNonan}.

Taking into account the explicit expressions of $D^{2,0}$ and $D^{0,2}$ from eqs. \p{DHanalB}, eqs. (\ref{DTranCo}a) and (\ref{DTranCo}b)
imply, respectively,  the following constraints on the original transformation parameters:
\bea
&& D^{2,0}\Lambda = 2i\left(\Lambda^{1,1}\theta^{1,-1} - \Lambda^{1,-1}\theta^{1,1} \right), \quad
D^{2,0}\Lambda^{1,1} = \Lambda^{2,0}\theta^{1,1}\,, \quad
D^{2,0}\Lambda^{1,-1} = \Lambda^{2,0}\theta^{1,-1}\,, \nn
&& D^{2,0}\Lambda^{-1,1} = \Lambda^{1,1} - \Lambda^{2,0}\theta^{-1,1}\,, \quad
D^{2,0}\Lambda^{2,0} =  D^{2,0}\Lambda^{0,2} = 0\,, \lb{D20constr} \\
&&  D^{2,0}\Lambda^{-1,-1} = \Lambda^{1,-1} - \Lambda^{2,0}\theta^{-1,-1} \lb{D20nonan}
\eea
and
\bea
&& D^{0,2}\Lambda = 0\,, \quad D^{0,2}\Lambda^{1,1} = \Lambda^{0,2}\theta^{1,1}\,, \quad
D^{0,2}\Lambda^{1,-1} = \Lambda^{1,1} -\Lambda^{0,2}\theta^{1,-1}\,, \nn
&& D^{0,2}\Lambda^{-1,1} = \Lambda^{0,2}\theta^{-1,1}\,, \quad
D^{0,2}\Lambda^{2,0} =  D^{0,2}\Lambda^{0,2} = 0\,, \lb{D02constr} \\
&&  D^{0,2}\Lambda^{-1,-1} = \Lambda^{-1,1} - \Lambda^{0,2}\theta^{-1,-1}\,. \lb{D02nonan}
\eea
The set of eqs. \p{D20constr} and \p{D02constr} fixes the structure of analytic parameters
$\Lambda^{1, \pm 1}, \Lambda^{-1,1}, \Lambda^{2,0}$ and $\Lambda^{0,2}$, whereas eqs. \p{D20nonan} and \p{D02nonan}
express the nonanalytic parameter $\Lambda^{-1,-1}$ in terms of the analytic ones. These analytic parameters
are given by the following expressions:
\bea
\Lambda &=& \omega + 2i\left(\theta^{1,1} \lambda^{ia}u^{-1}_iv^{-1}_a - \theta^{1,-1} \lambda^{ia}u^{-1}_iv^{1}_a \right) +
2i \theta^{1,1}\theta^{1,-1}\,\omega^{(ik)}u^{-1}_iu^{-1}_k\,, \lb{aa} \\
\Lambda^{1,1} &=& \lambda^{ia}u^{1}_iv^{1}_a + \theta^{1,1}\left[\frac{1}{2}\dot{\omega} + \omega^{(ik)}u^1_iu^{-1}_k +
\omega^{(ab)}v^1_av^{-1}_b \right] + i\theta^{-1,1}\theta^{1,-1} \left(\eta^{ia} -2\dot{\lambda}{}^{ia}\right) u^1_iv^1_a \nn
&& +\, i \theta^{1,1}\theta^{-1,1}\theta^{1,-1} \left(\phi - 2 \dot{\omega}{}^{(ab)}v^1_av^{-1}_b\right), \lb{ab} \\
\Lambda^{1,-1} &=& \lambda^{ia}u^{1}_iv^{-1}_a + \theta^{1,-1}\left[\frac{1}{2}\dot{\omega} + \omega^{(ik)}u^1_iu^{-1}_k -
\omega^{(ab)}v^1_av^{-1}_b \right] + \theta^{1,1}\omega^{(ab)} v^{-1}_av^{-1}_b\nn
&& +\,i \theta^{1,1}\theta^{1,-1}
\left(\eta^{ia} -2\dot{\lambda}{}^{ia}\right) u^{-1}_iv^{-1}_a\,,\lb{ac}\\
\Lambda^{-1,1} &=& \lambda^{ia}u^{-1}_iv^{1}_a + \theta^{-1,1}\left[\frac{1}{2}\dot{\omega} - \omega^{(ik)}u^1_iu^{-1}_k +
\omega^{(ab)}v^1_av^{-1}_b \right] + \theta^{1,1}\omega^{(ik)} u^{-1}_iu^{-1}_k \nn
&& - \, i\theta^{1,1}\theta^{-1,1} \eta^{ia}u^{-1}_iv^{-1}_a + 2i \theta^{1,-1}\theta^{-1,1}\dot{\lambda}^{ia}u^{-1}_iv^1_a
- 2i \theta^{1,1}\theta^{-1,1}\theta^{1,-1}\dot{\omega}{}^{(ik)}u^{-1}_iu^{-1}_k\,, \lb{ad}\\
\Lambda^{2,0} &=& \omega^{(ik)}u^1_iu^1_k - i\theta^{1,1}\eta^{ia}u^1_iv^{-1}_a + i\theta^{1,-1}\eta^{ia}u^1_iv^{1}_a \nn
&&- \,i\theta^{1,1}\theta^{1,-1}
\left(\phi + \ddot{\omega} + 2\dot{\omega}{}^{(ik)}u^1_iu^{-1}_k\right), \lb{ae}\\
\Lambda^{0,2} &=& \omega^{(ab)}v^1_av^1_b + i\theta^{1,1}\left(\eta^{ia} -2\dot\lambda{}^{ia}\right) u^{-1}_iv^{1}_a
- i\theta^{-1,1}\left(\eta^{ia} - 2\dot\lambda{}^{ia}\right)u^1_iv^{1}_a \nn
&&+ \,i\theta^{1,1}\theta^{-1,1}
\left(\phi - 2\dot{\omega}{}^{(ab)}v^1_a v^{-1}_b\right)
+  2i\theta^{1,-1}\theta^{-1,1}\dot{\omega}{}^{(ab)}v^1_a v^{1}_b \nn
&& +\,2\theta^{1,1}\theta^{-1,1}\theta^{1,-1}\left(\dot{\eta}{}^{ia} - 2\ddot{\lambda}{}^{ia} \right)u^{-1}_iv^1_a\,. \lb{af}
\eea
Thus we are left with the set of $(8 + 8)$ real functions as the essential parameters:
\bea
&& \underline{\mbox{bosonic}}: \qquad \omega(t)\,, \;\omega^{(ik)}(t)\,, \;\omega^{(ab)}(t), \;\phi(t)\,, \nn
&& \underline{\mbox{fermionic}}: \qquad \lambda^{ia}(t)\,, \; \eta^{ia}(t)\,. \lb{Param}
\eea
These functions parametrize a classical centerless version of the so-called large infinite-dimensional
${\cal N}{=}4$ $\rm{SO}(4)\times \rm{U}(1)$ superconformal group \cite{Adem,Sev,IKLq}.
The bosonic functions $\omega(t), \phi(t)$  and $\omega^{(ik)}(t),
\omega^{(ab)}(t)\,$, in their $t$ -expansion, collect the parameters of the Virasoro, of the $\rm{U}(1)\,$,
and of the $\rm{SU}(2)\times \rm{SU}(2)$
($\sim \rm{SO}(4)$) Kac-Moody
transformations,\footnote{Strictly speaking, the full set of the Kac-Moody $\rm{U}(1)$ transformations is collected in the ``prepotential''
$\varphi(t)$ of the function $\phi \sim \dot{\varphi}$ (see discussion on p. 11).}
while the fermionic functions $\lambda^{ia}(t)$ and $\eta^{ia}(t)$ include the fermionic
parameters associated with the ``canonical'' and ``noncanonical'' superconformal generators, respectively. All bosonic
symmetries are contained in the closure of the fermionic variations, which looks like
\bea
[\delta_{(\lambda)}, \delta_{(\lambda)}] \;\sim \; \delta_{(\omega)} + \delta_{(\omega^{ab})}\,, \quad
[\delta_{(\lambda)}, \delta_{(\eta)}] \;\sim \; \delta_{(\omega)} + \delta_{(\phi)} + \delta_{(\omega^{ab})} + \delta_{(\omega^{ik})}\,, \quad
[\delta_{(\eta)}, \delta_{(\eta)}] = 0\,.
\eea
The expression for the nonanalytic superparameter function $\Lambda^{-1,-1}$ in terms of the independent parameter functions \p{Param}
can be directly found by solving Eqs. \p{D20nonan} and \p{D02nonan}.

The superconformal transformations \p{aa} - \p{af} preserve the three-theta analyticity but, by no means, the
two-theta analyticities. The most characteristic feature of this realization  is that both sets of harmonics,
i.e. $u^{\pm 1}_i$ and $v^{\pm 1}_a$,
undergo the local (Kac-Moody) SU(2) transformations in their doublet indices with the parameters $\omega^{(ik)}$ and
$\omega^{(ab)}$, respectively. However, there exist two different realizations of the same large ${\cal N}{=}4$ superconformal group
preserving the two-theta analyticities \p{Asubs+} and \p{Asubs-}. Because these analytic subspaces are ``mirror images''
of each other, it will be enough to consider, for instance, the case of \p{Asubs+}.

We begin with the two-theta analyticity-preserving counterparts of the transformations \p{TranCooan} - \p{TranNonan}
\footnote{Tildes used below should
be distinguished from those in the generalized $\;\widetilde{\;\;\;}\;$ conjugation.}:
\bea
&& \delta t_+ =
\tilde{\Lambda}_+(\zeta_+, u, v)\,, \;\; \delta \theta^{1,1} =
\tilde{\Lambda}^{1,1}(\zeta_+, u, v)\,, \;\;
\delta \theta^{1,-1} = \tilde{\Lambda}^{1,-1}(\zeta_+, u, v)\,, \lb{TranCooan2} \\
&&\delta u^{1}_i = \tilde{\Lambda}^{2,0}(\zeta_+, u, v)u^{-1}_i\,, \;\delta u^{-1}_i = 0\,,
\quad \delta v^{1}_a = \tilde{\Lambda}^{0,2}(\zeta_+, u, v)v^{-1}_a\,, \;\delta v^{-1}_a = 0\,,\lb{TranHarm2} \\
&&\delta \theta^{-1,-1} = \tilde{\Lambda}^{-1,-1}(\zeta_+, \theta^{-1,1}, \theta^{-1,-1}, u, v)\,, \quad
\delta \theta^{-1,1} = \tilde{\Lambda}^{-1,1}(\zeta_+, \theta^{-1,1}, \theta^{-1,-1}, u, v)\,, \lb{TranNonan2}
\eea
and, once again, look for their closed subset consisting of those transformations which preserve
the form of harmonic derivatives $D^{2,0}, D^{0.2}$:
\be
(\mbox{a})\;\; \delta D^{2,0} = -\tilde{\Lambda}^{2,0}D^0_u\,, \quad (\mbox{b})\;\;\delta D^{0,2} = -\tilde{\Lambda}^{0,2}D^0_v\,. \lb{DTranCo2}
\ee
The resulting equations for the superparameters mimic Eqs. \p{D20constr} - \p{D02nonan}, namely,
\bea
&& D^{2,0}\tilde{\Lambda}_+ = 2i\left(\tilde{\Lambda}{}^{1,1}\theta^{1,-1} - \tilde{\Lambda}{}^{1,-1}\theta^{1,1} \right), \quad
D^{2,0}\tilde{\Lambda}{}^{1,1} = \tilde{\Lambda}{}^{2,0}\theta^{1,1}\,,
D^{2,0}\tilde{\Lambda}{}^{1,-1} = \tilde\Lambda{}^{2,0}\theta^{1,-1}\,, \nn
&& D^{2,0}\tilde\Lambda{}^{2,0} =  D^{2,0}\tilde\Lambda{}^{0,2} = 0\,, \lb{D20constr2} \\
&& D^{2,0}\tilde\Lambda{}^{-1,1} = \tilde\Lambda{}^{1,1} - \tilde\Lambda{}^{2,0}\theta^{-1,1}\,, \quad
D^{2,0}\tilde\Lambda{}^{-1,-1} = \tilde\Lambda{}^{1,-1} - \tilde\Lambda{}^{2,0}\theta^{-1,-1} \lb{D20nonan2}
\eea
and
\bea
&& D^{0,2}\tilde\Lambda_+ = 0\,, \quad D^{0,2}\tilde\Lambda{}^{1,1} = \tilde\Lambda{}^{0,2}\theta^{1,1}\,, \quad
D^{0,2}\tilde\Lambda{}^{1,-1} = \tilde\Lambda{}^{1,1} -\tilde\Lambda{}^{0,2}\theta^{1,-1}\,, \nn
&& D^{0,2}\tilde\Lambda{}^{2,0} =  D^{0,2}\tilde\Lambda{}^{0,2} = 0\,, \lb{D02constr2} \\
&& D^{0,2}\tilde\Lambda{}^{-1,1} = \tilde\Lambda{}^{0,2}\theta^{-1,1}\,, \quad
D^{0,2}\tilde\Lambda{}^{-1,-1} = \tilde\Lambda{}^{-1,1} - \tilde\Lambda{}^{0,2}\theta^{-1,-1}\,. \lb{D02nonan2}
\eea
However, there is an essential difference from the previous case (besides the more restrictive two-theta analyticity
of $\tilde{\Lambda}_+, \tilde\Lambda^{1, -1}, \tilde\Lambda^{2,0}, \tilde{\Lambda}^{0,2}$). It consists in that superparameter
$\tilde\Lambda^{-1,1}$ is now a function defined on the whole bi-HSS, while its analogue $\Lambda^{-1,1}$ has been required
to live on the three-theta analytic subspace. Because of this, Eqs. \p{D20constr2} - \p{D02nonan2}
cannot be obtained as some truncation of Eqs. \p{D20constr} - \p{D02nonan}, as one might naively think.
Therefore, their general solution is by no means a particular case of \p{aa} - \p{af}. It can be written as
\bea
\tilde\Lambda_+ &=& \tilde\omega + 2i\,(\theta^{1,1} \tilde\lambda{}^{ia}u^{-1}_iv^{-1}_a -
\theta^{1,-1} \tilde\lambda{}^{ia}u^{-1}_iv^{1}_a ) +
2i \theta^{1,1}\theta^{1,-1}\,\tilde{\omega}{}^{(ik)}u^{-1}_iu^{-1}_k\,, \lb{Aa} \\
\tilde{\Lambda}{}^{1,1} &=& \tilde\lambda{}^{ia}u^{1}_iv^{1}_a + \theta^{1,1}\left[\frac{1}{2}\dot{\tilde\omega}
+ \tilde\omega{}^{(ik)}u^1_iu^{-1}_k +
(\hat\tau{}^{(ab)} + \tilde{\omega}{}^{(ab)})v^1_av^{-1}_b \right] \nn
&& -\, \theta^{1,-1}(\hat\tau{}^{(ab)}
+ \tilde{\omega}{}^{(ab)})v^1_av^{1}_b + i \theta^{1,1}\theta^{1,-1}(\tilde{\eta}{}^{ia}
- 2 \dot{\tilde\lambda}{}^{ia})u^{-1}_iv^{1}_a, \lb{Ab} \\
\tilde\Lambda{}^{1,-1} &=& \tilde\lambda{}^{ia}u^{1}_iv^{-1}_a + \theta^{1,-1}\left[\frac{1}{2}\dot{\tilde\omega}
+ \tilde\omega{}^{(ik)}u^1_iu^{-1}_k -
(\hat\tau{}^{(ab)} + \tilde\omega{}^{(ab)})v^1_av^{-1}_b \right] \nn
&& +\, \theta^{1,1}(\hat\tau{}^{(ab)} + \tilde{\omega}{}^{(ab)}) v^{-1}_av^{-1}_b + i \theta^{1,1}\theta^{1,-1}
(\tilde\eta{}^{ia} -2\dot{\tilde\lambda}{}^{ia}) u^{-1}_iv^{-1}_a\,,\lb{Ac} \\
\tilde{\Lambda}^{2,0} &=& \tilde{\omega}^{(ik)}u^1_iu^1_k - i\theta^{1,1}\tilde{\eta}{}^{ia}u^1_iv^{-1}_a
+ i\theta^{1,-1}\tilde{\eta}{}^{ia}u^1_iv^{1}_a
- i\theta^{1,1}\theta^{1,-1}
(\tilde\phi + \ddot{\tilde\omega} + 2\dot{\tilde\omega}{}^{(ik)}u^1_iu^{-1}_k), \lb{Ae}\\
\tilde\Lambda{}^{0,2} &=& \hat{\tau}^{(ab)}v^1_av^1_b\,.  \lb{Af}
\eea

The functions
\bea
&& \underline{\mbox{bosonic}}: \qquad \tilde\omega(t)\,, \;\tilde\omega{}^{(ik)}(t)\,, \;\tilde\omega{}^{(ab)}(t), \;\tilde\phi(t)\,, \nn
&& \underline{\mbox{fermionic}}: \qquad \tilde\lambda{}^{ia}(t)\,, \; \tilde\eta{}^{ia}(t)\lb{Param2}
\eea
also parametrize the classical large ${\cal N}{=}4$ $\rm{SO}(4)\times \rm{U}(1)$ superconformal group. The closure of transformations
\p{TranCooan2} - \p{DTranCo2} has the same structure in terms of these parameters as that of \p{TranCooan} - \p{DTranCo}
in terms of the parameters \p{Param}.
The additional SU(2) parameters $\hat{\tau}{}^{(ab)}$ are constant, and the corresponding transformations
form a semidirect product with the superconformal group. Thus, a distinguishing feature of the two-theta analyticity-preserving
realization \p{Aa} - \p{Ae} of the large superconformal group is that only one ``conformal'' local (Kac-Moody)
SU(2) symmetry has a nontrivial action on the harmonic variables: this is the group with parameters $\tilde\omega{}^{(ik)}(t)$
acting on harmonics $u^1_i$. Another SU(2) factor of the full SO(4) Kac-Moody subgroup (the one with parameters
$\tilde\omega{}^{(ab)}(t)$) does not affect the harmonics $v^{\pm 1}_a$ and acts only on the Grassmann
coordinates. In the mirror realization which preserves the alternative two-theta analytic subspace \p{Asubs-}, the roles of these two
SU(2) Kac-Moody factors are exchanged: the group with parameters $\tilde\omega{}^{(ab)}(t)$ has a nontrivial
action on harmonics $v^1_a$ while the group with parameters $\tilde\omega{}^{(ik)}(t)$ affects only the Grassmann coordinates.
Let us remark that the mirror realization can be obtained from \p{Aa} - \p{Ae} by the following mnemonic rules
\bea
&&\mbox{(i)} \;\; t_+ \,\rightarrow\, t_-; \qquad \mbox{(ii)} \;\; \theta^{\pm 1, \pm 1} \,\leftrightarrow\,
\theta^{\pm 1, \pm 1}\,, \;\; \theta^{\pm 1, \mp 1} \,\leftrightarrow\,
\theta^{\mp 1, \pm 1}; \nn
&& \mbox{(iii)} \;\; i \, \leftrightarrow \, a\,, \;\; (m, n) \, \rightarrow \, (n, m)\,, \;\; u^{\pm 1}_i \,\leftrightarrow \,
v^{\pm 1}_a\,. \lb{mnem}
\eea
Here, $(m, n)$, as before, denotes the harmonic $\rm{U}(1) \times \rm{U}(1)$ charges of different quantities. The three-theta
realization \p{aa} - \p{af} is closed under these changes.

It is worthwhile to note two things. First, explicit expressions for nonanalytic superparameters $\tilde{\Lambda}^{-1\pm 1}$
in terms of the parameters \p{Param2}
can be found from Eqs. \p{D20nonan2} and \p{D02nonan2}. Second, one can pass to an analytic superspace with one
set of harmonics $u^\pm_i$ by substituting $\theta^{1, \pm 1} = \theta^{+ a}v^{\pm 1}_a$. The coordinate set
$(t_+, \theta^{+ a}, u^{\pm 1}_i \equiv u^\pm_i)$ parametrizes  an analytic subspace of the standard ${\cal N}{=}4, d{=}1$ harmonic
superspace \cite{IL}. Since harmonics $v^{\pm 1}_a$ are inert with respect to realization \p{Aa} - \p{Ae}, the latter can be equivalently
rewritten in terms of this coordinate set just by taking off the harmonics $v^{\pm 1}_a$ from the left-hand and from the right-hand sides of
\p{Aa} - \p{Ae}. Realization of the large ${\cal N}{=}4$ superconformal group preserving this analytic ${\cal N}{=}4$
superspace with one set of harmonic variables was first found in \cite{DS}, by also requiring the analyticity-preserving
harmonic derivative ($D^{++}$ in this case) to retain its flat form. In the previously considered case
of the three-theta analytic superspace realization \p{aa} - \p{af}, one cannot take off either harmonics $v^{\pm 1}_a$ or
$u^{\pm 1}_i$, since both harmonic sets are nontrivially transformed by the superconformal group. The realization \p{aa} - \p{af}
is new, and is inherent just to the three-theta ${\cal N}{=}4, d{=}1$ analyticity which is a specific feature
of the bi-HSS.

It is known that the large ${\cal N}{=}4$ superconformal group includes two ``small'' ${\cal N}{=}4$ SU(2)
superconformal groups as subgroups. They contain  only the Virasoro and the SU(2) Kac-Moody groups
in their bosonic sector, accompanied by one set
of the fermionic superconformal generators. In the realization \p{aa} - \p{af}, transformations of these two small superconformal
groups correspond to the following alternative truncations of the set of parameter functions:
\bea
\eta^{ia}(t) = \omega^{(ik)}(t) = 0\,, \; \phi(t) = -\ddot{\omega}(t)  \, &\Rightarrow & \, {\cal N}{=}4\; {\rm SU}(2)\;I: \;
\omega(t)\,, \, \lambda^{ia}(t)\,, \,\omega^{(ab)}(t)\,, \lb{SCAI} \\
\eta^{ia}(t) = 2 \dot{\lambda}{}^{ia}(t)\,, \; \omega^{(ab)}(t) = \phi(t) = 0  \, &\Rightarrow & \, {\cal N}{=}4\; {\rm SU}(2)\;II: \;
\omega(t)\,, \, \lambda^{ia}(t)\,, \, \omega^{(ik)}(t)\,. \lb{SCAII}
\eea
One can directly check that transformations with superparameters
\p{aa} - \p{af} involving the relevant smaller sets of the parameter-functions are still closed under the Lie brackets and form two
centerless ${\cal N}{=}4$ SU(2) superconformal groups. The same truncations, in which all parameters are changed to those with tilde,
single out two isomorphic ${\cal N}{=}4$ SU(2) superconformal groups in the realization \p{Aa} - \p{Ae}.

Finally, we would like to mention that the full set of parameters of the $\rm{U}(1)$ Kac-Moody subgroups in the realizations
\p{aa} - \p{af} and \p{Aa} - \p{Ae} is actually reproduced after passing to the analytic ``prepotentials'' for the
superparameters $\Lambda^{2,0}, \Lambda^{0,2}\,$, and $\tilde{\Lambda}^{2,0}$ as
\be
\Lambda^{2,0} = D^{2,0}\Lambda_L\,, \quad \Lambda^{0,2} = D^{0,2}\Lambda_R\,, \quad \tilde{\Lambda}^{2,0} = D^{2,0}\tilde{\Lambda}_L\,.
\ee
Here,
\bea
&&\tilde{\Lambda}_L = \tilde\varphi + \tilde{\omega}{}^{(ik)}u^1_iu^{-1}_k - i\theta^{1,1}\tilde{\eta}{}^{ia}u^{-1}_iv^{-1}_a
+ i\theta^{1,-1}\tilde{\eta}{}^{ia}u^{-1}_iv^{1}_a
- 2i\theta^{1,1}\theta^{1,-1}\dot{\tilde\omega}{}^{(ik)}u^{-1}_iu^{-1}_k\,, \nn
&&\Lambda_L = \varphi + \omega{}^{(ik)}u^1_iu^{-1}_k - i\theta^{1,1}\eta{}^{ia}u^{-1}_iv^{-1}_a
+ i\theta^{1,-1}\eta{}^{ia}u^{-1}_iv^{1}_a
- 2i\theta^{1,1}\theta^{1,-1}\dot{\omega}{}^{(ik)}u^{-1}_iu^{-1}_k\,, \nn
&&\Lambda_R = -\varphi -\frac{1}{2}\dot\omega + \omega^{(ab)}v^1_av^{-1}_b +i\theta^{1,1}\left(\eta^{ia}
-2\dot\lambda{}^{ia}\right) u^{-1}_iv^{-1}_a
- i\theta^{-1,1}\left(\eta^{ia} - 2\dot\lambda{}^{ia}\right)u^1_iv^{-1}_a \nn
&& -\,2i\theta^{1,1}\theta^{-1,1}\dot{\omega}{}^{(ab)}v^{-1}_a v^{-1}_b
+  i\theta^{1,-1}\theta^{-1,1}\left[\phi + 2\dot{\omega}{}^{(ab)}v^1_a v^{-1}_b\right] \nn
&& +\, 2\theta^{1,1}\theta^{-1,1}\theta^{1,-1}\left(\dot{\eta}{}^{ia} - 2\ddot{\lambda}{}^{ia} \right)u^{-1}_iv^{-1}_a\,, \;\;
\tilde{\phi} = -(2\dot{\tilde\varphi} + \ddot{\tilde\omega})\,, \, \phi = -(2\dot{\varphi} + \ddot{\omega})\,, \nn
&& D^{0,2}\tilde{\Lambda}_L =D^{0,2}\Lambda_L = D^{2,0}\Lambda_R = 0\,. \lb{lambda}
\eea
The new dimensionless parameters $\tilde\varphi(t), \varphi(t)$ expanded in the Taylor series with
respect to $t$ produce a full set of the Kac-Moody $\rm{U}(1)$ parameters, including the rigid $\rm{U}(1)$ symmetry
parameters $\tilde\varphi(0), \varphi(0)$, in full agreement with the structure of the ${\cal N}{=}4$ $\rm{SO}(4)\times \rm{U}(1)$
superconformal algebra \cite{Sev,IKLq}.  The parameters $\tilde\varphi(0)$ and $\varphi(0)$ do not show up in the above realizations
on the superspace coordinates, but can appear in realizations on superfields, with $\tilde{\Lambda}_L, \Lambda_L$ and $\Lambda_R$
as weight factors.\footnote{See, e.g., \cite{IS,BI} for an analogous phenomenon in $d{=}2$ sigma models.}

\subsection{Finite-dimensional superconformal group $D(2,1;\alpha)$}

It is well known that the maximal finite-dimensional subgroup of the large ${\cal N}{=}4$ $\rm{SO}(4)\times \rm{U}(1)$  superconformal group is
the supergroup $D(2,1;\alpha)$ \cite{Sorba}, whereas, in the small ${\cal N}{=}4$ superconformal group, the same role is played by
the supergroup $\rm{SU}(1,1|2)$ which can be treated as a particular case of $D(2,1;\alpha)$ with $\alpha =-1$
or $\alpha =0\,$.\footnote{Actually,
these cases yield a semidirect product of $\rm{SU}(1,1|2)$ with an extra SU(2) which does appear in the anticommutators of the fermionic
generators. Generically, our choice of parameter $\alpha$ is such that two $\rm{SU}(2) \subset D(2,1;\alpha)$ enter the right-hand sides
of the anticommutator of the fermionic generators with the coefficients $\alpha$ and $-(1 +\alpha)$, like in \cite{Sorba}.}
In our case the subgroups $D(2,1;\alpha)$ of the realizations \p{aa} - \p{af} and \p{Aa} - \p{Ae} are extracted in the following
unique way.

As an example, we consider the realization \p{aa} - \p{af}. First, we restrict the infinite-dimensional conformal
group associated with parameter $\omega(t)$ down to the
finite-dimensional $d{=}1$ conformal group $\rm{SO}(2,1) \sim \rm{SU}(1,1)$ by imposing the constraint
\be
\dddot{\omega} = 0 \;\; \Rightarrow \;\;\omega = \omega_0 + t\omega_1 + t^2\omega_2\,.\lb{confcond}
\ee
Here, $\omega_0\,,\, \omega_1$ and $\omega_2$ are constant parameters of the $d{=}1$ translations,
dilatations and conformal boosts, respectively.
Then we seek the most general constraints on the remaining parameter functions, such
that the Lie brackets of different transformations
are compatible with \p{confcond}. They are
\be
\ddot{\lambda}{}^{ia} = 0 \;\;\Rightarrow \;\; \lambda{}^{ia} = \varepsilon^{ia} + t \beta^{ia}\,,\lb{sconfcond}
\ee
where $\varepsilon^{ia}, \beta^{ia}$ are constant parameters of the ${\cal N}{=}4, d{=}1$ Poincar\'e and conformal
supersymmetries, and also
\bea
&& \eta^{ia} = -2\alpha\, \dot{\lambda}{}^{ia} = -2\alpha\, \beta^{ia}\,, \quad \phi = -(1 +\alpha)\,\ddot{\omega} =
-2(1 +\alpha)\,\omega_2\,, \quad \varphi = \frac{\alpha}{2}\,\dot\omega\,, \nn
&& \omega^{(ik)} = -\alpha\, \tau^{(ik)}\,, \;\omega^{(ab)} = (1 +\alpha)\,\tau^{(ab)}\,, \quad  \dot{\tau}{}^{(ik)}
= \dot{\tau}{}^{(ab)} = 0\,.\lb{sconfcond2}
\eea
Here, $\alpha$ is an arbitrary real parameter. It can be checked that the ${\cal N}{=}4$ superconformal transformations
form just the supergroup $D(2,1;\alpha)$, including its extreme $\alpha =0$ and $\alpha = -1$ $\rm{SU}(1,1|2)$
cases, as well as the case of $\alpha = -\frac{1}{2}\,$, which yields the supergroup $\rm{OSp}(4|2)\,$.\footnote{There are equivalent
choices of $\alpha$ which give rise to the isomorphic superalgebras \cite{Sorba}.} Conditions singling out the $D(2,1;\alpha)$
transformations in set \p{Aa} - \p{Ae} are the same as in \p{confcond} - \p{sconfcond2} (with ``tildes''
on the relevant parameter functions).

Like in the case of infinite-dimensional ${\cal N}{=}4$ superconformal groups, all bosonic symmetries
of $D(2,1;\alpha)$ are contained
in the closure of its transformations associated with odd parameters \p{sconfcond}. For further purposes, we present relevant
pieces of both sets \p{aa} - \p{af} and \p{Aa} - \p{Ae} adapted to the finite-dimensional case explicitly,
including nonanalytic superfunctions.
We shall use the same notation for the parameters of $D(2,1;\alpha)$, despite the fact that the form of the transformations
in these two cases is different.

In \p{aa} - \p{af}, we have:
\bea
&&\Lambda \,\Rightarrow \, 2i\theta^{1,1}\lambda^{-1,-1}  - 2i\theta^{1,-1}\lambda^{-1,1}\,, \;
\Lambda^{1,1}\, \Rightarrow \,\lambda^{1,1} + 2i(1+\alpha)\,\theta^{1,-1}\theta^{-1,1} \dot{\lambda}{}^{1,1}\,, \nn
&& \Lambda^{1,-1}\, \Rightarrow \,\lambda^{1,-1} - 2i(1+\alpha)\,\theta^{1,1}\theta^{1,-1} \dot{\lambda}{}^{-1,-1}\,, \nn
&& \Lambda^{-1,1}\, \Rightarrow \,\lambda^{-1,1} + 2i\alpha\,\theta^{1,1}\theta^{-1,1} \dot{\lambda}{}^{-1,-1}
+ 2i\,\theta^{1,-1}\theta^{-1,1} \dot{\lambda}{}^{-1,1}\,, \nn
&& \Lambda^{-1,-1}\, \Rightarrow \, \lambda^{-1,-1} - 2i\,\theta^{1,1}\theta^{-1,-1} \dot{\lambda}{}^{-1,-1}
- 2i\alpha\,\theta^{1,-1}\theta^{-1,-1} \dot{\lambda}{}^{-1,1} \nn
&&{\quad \quad \quad \;} +\, 2i(1+\alpha)\,\theta^{1,-1}\theta^{-1,1} \dot{\lambda}{}^{-1,-1}
+ 2i(1+\alpha)\,\theta^{-1,1}\theta^{-1,-1} \dot{\lambda}{}^{1,-1}\,,\nn
&& \Lambda^{2,0}\, \Rightarrow \, 2i\alpha\left(\theta^{1,1}\dot{\lambda}{}^{1,-1} - \theta^{1,-1}\dot{\lambda}{}^{1,1}\right),
\, \Lambda_L \,\Rightarrow \, 2i\alpha\left(\theta^{1,1}\dot{\lambda}{}^{-1,-1} - \theta^{1,-1}\dot{\lambda}{}^{-1,1}\right),\nn
&&\Lambda^{0,2}\, \Rightarrow \,- 2i(1+\alpha)\left(\theta^{1,1}\dot{\lambda}{}^{-1,1}
- \theta^{-1,1}\dot{\lambda}{}^{1,1}\right), \nn
&&\Lambda_R \,\Rightarrow \,- 2i(1+\alpha)\left(\theta^{1,1}\dot{\lambda}{}^{-1,-1}
- \theta^{-1,1}\dot{\lambda}{}^{1,-1}\right), \lb{alphaI}
\eea
where $\lambda^{1,1} = \lambda^{ia}u^1_iv^1_a\,$, etc.

In \p{Aa} - \p{Ae}, we have:
\bea
&&\tilde\Lambda_+ \,\Rightarrow \, 2i\theta^{1,1}\lambda^{-1,-1}  - 2i\theta^{1,-1}\lambda^{-1,1}\,, \;
\tilde\Lambda{}^{1,1}\, \Rightarrow \,\lambda^{1,1} - 2i(1+\alpha)\,\theta^{1,1}\theta^{1,-1} \dot{\lambda}{}^{-1,1}\,, \nn
&& \tilde\Lambda{}^{1,-1}\, \Rightarrow \, \lambda^{1,-1} - 2i(1+\alpha)\,\theta^{1,1}\theta^{1,-1} \dot{\lambda}{}^{-1,-1}\,, \nn
&& \tilde\Lambda{}^{-1,1}\, \Rightarrow \, \lambda^{-1,1} + 2i\alpha\,\theta^{1,1}\theta^{-1,1} \dot{\lambda}{}^{-1,-1}
+ 2i\,\theta^{1,-1}\theta^{-1,1} \dot{\lambda}{}^{-1,1}\,, \nn
&&{\quad \quad \quad \;} -\, 2i(1+\alpha)\,\theta^{1,1}\theta^{-1,-1} \dot{\lambda}{}^{-1,1}
+ 2i(1+\alpha)\,\theta^{-1,1}\theta^{-1,-1} \dot{\lambda}{}^{1,1}\,,\nn
&& \tilde\Lambda{}^{-1,-1}\, \Rightarrow \,\lambda^{-1,-1} - 2i\,\theta^{1,1}\theta^{-1,-1} \dot{\lambda}{}^{-1,-1}
- 2i\alpha\,\theta^{1,-1}\theta^{-1,-1} \dot{\lambda}{}^{-1,1} \nn
&&{\quad \quad \quad \;} +\, 2i(1+\alpha)\,\theta^{1,-1}\theta^{-1,1} \dot{\lambda}{}^{-1,-1}
+ 2i(1+\alpha)\,\theta^{-1,1}\theta^{-1,-1} \dot{\lambda}{}^{1,-1}\,,\nn
&& \tilde\Lambda{}^{2,0}\, \Rightarrow \,2i\alpha\left(\theta^{1,1}\dot{\lambda}{}^{1,-1}
- \theta^{1,-1}\dot{\lambda}{}^{1,1}\right), \quad \tilde\Lambda^{0,2}\, \Rightarrow \,0\,, \nn
&& \tilde{\Lambda}_L\,\Rightarrow \, 2i\alpha\left(\theta^{1,1}\dot{\lambda}{}^{-1,-1}
- \theta^{1,-1}\dot{\lambda}{}^{-1,1}\right).
\lb{alphaII}
\eea
Notice that the mirror realization of the same superconformal group $D(2,1;\alpha)$ preserving
the analytic superspace \p{Asubs-}
can be obtained from \p{alphaII} by changes \p{mnem} together with the replacement
\be
\alpha \;\rightarrow \; -(1 + \alpha)\,. \lb{addmnem}
\ee
Under these changes realization \p{alphaI} is ``self-conjugate.''

It will be important to specify how the integration measures \p{FullMea}, \p{AMea+} and \p{AMea3} are transformed
under these $D(2,1;\alpha)$
realizations. Using the general formula
\be
\delta \hat\mu =
\hat\mu\left(\partial_{t_+}\delta t_+ + \partial_{u^1_i}\delta u^1_i + \partial_{v^1_a}\delta v^1_a
- \sum \partial_{n,m}\delta \theta^{n,m}\right),
\ee
where $\hat{\mu}$ stands for any measure \p{FullMea} - \p{AMea3}, we have found that, under the conformal supersymmetry, the measures
transform as
\bea
&&\delta \mu = 4i\mu \left[ (\,\theta^{-1,1}\dot{\lambda}{}^{1,-1}- \theta^{1,1}\dot{\lambda}{}^{-1,-1}\,) +
\alpha\, (\,\theta^{-1,1}\dot{\lambda}{}^{1,-1}- \theta^{1,-1}\dot{\lambda}{}^{-1,1}\,) \right], \lb{TranMeI} \\
&&\delta \mu_{A3}^{(-1,-1)}  = 2i\mu_{A3}^{(-1,-1)}\left[ (\,\theta^{-1,1}\dot{\lambda}{}^{1,-1}- \theta^{1,1}\dot{\lambda}{}^{-1,-1}\,) +
\alpha\, (\,\theta^{-1,1}\dot{\lambda}{}^{1,-1}- \theta^{1,-1}\dot{\lambda}{}^{-1, 1}\,) \right] \lb{TranMeaI}
\eea
for realization \p{alphaI} and as
\bea
&&\delta \mu = 2i \mu\left[ (1 -\alpha)\,(\,\theta^{1,-1}\dot{\lambda}{}^{-1,1} - \theta^{1,1}\dot{\lambda}{}^{-1,-1}\,) +
(1+\alpha)\, (\,\theta^{-1,1}\dot{\lambda}{}^{1,-1}- \theta^{-1,-1}\dot{\lambda}{}^{1,1}\,) \right], \lb{TranMeII} \\
&&\delta \mu_{A+}^{(-2,0)}  = 0 \lb{TranMeaII}
\eea
for realization \p{alphaII}. A difference in the transformations of the full integration measure $\mu$ is related to the property
that harmonics $v^1_a$ undergo a nontrivial transformation in the first case and are inert in the second case.
Transformations \p{TranMeI}, \p{TranMeaI} are not affected by
mirror changes \p{mnem}, \p{addmnem},  while the variation \p{TranMeII} is converted into
\bea
\delta \mu = 2i\mu \left[ (2 +\alpha)\,(\,\theta^{-1,1}\dot{\lambda}{}^{1,-1} - \theta^{1,1}\dot{\lambda}{}^{-1,-1}\,)
-\alpha\, (\,\theta^{1,-1}\dot{\lambda}{}^{-1,1}- \theta^{-1,-1}\dot{\lambda}{}^{1,1}\,) \right]. \lb{TranMeIII}
\eea
This reflects the fact that the realizations of $D(2,1;\alpha)$ preserving the analytic subspaces \p{Asubs+} and \p{Asubs-}
are essentially different: they cannot be related to each other by any redefinition of the superspace coordinates.

It is worth noting that in the case of the three-theta analyticity-preserving realization, the full $D(2,1;\alpha)$
transformations of measures $\mu_{A3}$ and $\mu$ can be written as
\be
\delta \mu_{A3}^{(-1,-1)} = \mu_{A3}^{(-1,-1)}\left(\Lambda_L + \Lambda_R\right), \quad \delta \mu =
2\mu\left(\Lambda_L + \Lambda_R\right),
\ee
where $\Lambda_L$ and $\Lambda_R$ were defined in \p{lambda} (with conditions \p{confcond} - \p{sconfcond2}
taken into account in the $D(2,1;\alpha)$ case).

We need transformation properties of harmonic derivatives
$D^{-2,0}$ and $D^{0,-2}$ under the superconformal boost generators
of $D(2,1;\alpha)\,$. For realizations \p{alphaI}, these
derivatives transform as
\bea &&
\delta D^{-2, 0} = -2i\alpha
\left[D^{-2,0}\left(\theta^{1,1}\dot\lambda{}^{1,-1} -
\theta^{1,-1}\dot{\lambda}{}^{1,1} \right)\right] D^{-2, 0}\,, \nn
&& \delta D^{0,-2} = 2i(1+\alpha)
\left[D^{0,-2}\left(\theta^{1,1}\dot\lambda{}^{-1,1} -
\theta^{-1,1}\dot{\lambda}{}^{1,1} \right)\right] D^{0, -2}\,,
\lb{D-I}
\eea
and, for \p{alphaII}, as
\bea && \delta D^{-2, 0} =
-2i\alpha \left[D^{-2,0}\left(\theta^{1,1}\dot\lambda{}^{1,-1} -
\theta^{1,-1}\dot{\lambda}{}^{1,1} \right)\right] D^{-2, 0}\,, \quad
\delta D^{0,-2} = 0\,. \lb{D-II}
\eea

\setcounter{equation}{0}
\section{${\cal N}{=}4$ supermultiplets in the biharmonic superspace}
Various ${\cal N}{=}4$ supermultiplets with a finite number of component fields admit
a simple concise description in the bi-HSS. Later on, we shall list these multiplets and some of their
superfield actions. In a few instructive cases, we shall also discuss their properties with respect to the
${\cal N}{=}4$ superconformal group $D(2,1;\alpha)\,$.

\subsection{Multiplets ${\bf (4,4,0)}$}
Multiplet ${\bf (4,4,0)}$ exists in two basic complementary forms which differ in the SU(2)
assignment of component fields.  In the bi-HSS they are represented by the superfields $q^{(1,0)\underline{A}}$ and
$q^{(0,1)A}$ ($\underline{A}, A = 1,2$) subjected to the following analyticity conditions and harmonic constraints:
\bea
\mbox{(a)} \;D^{1,1} q^{(1,0)\underline{A}} = D^{1,-1} q^{(1,0)\underline{A}} = 0\,, \quad
\mbox{(b)} \; D^{2,0}q^{(1,0)\underline{A}} =  D^{0, 2,}q^{(1,0)\underline{A}} = 0\,, \lb{cq1}
\eea
and
\bea
\mbox{(a)} \;D^{1,1} q^{(0,1) A} = D^{-1,1} q^{(0,1)A} = 0\,, \quad
\mbox{(b)} \; D^{2,0}q^{(0,1)A} =  D^{0, 2,}q^{(0,1)A} = 0\,. \lb{cq2}
\eea
The extra doublet indices $\underline{A}$ and $A$ refer to some extra SU(2) groups commuting with the ${\cal N}{=}4$ supersymmetry
(the so-called Pauli-G\"ursey ({\it PG}) SU(2) groups \cite{HSS}). The above superfields  are assumed
to satisfy the reality conditions
\be
\widetilde{(q^{(1,0)\underline{A}})} = \epsilon_{\underline{A}\underline{B}}q^{(1,0)\underline{B}}\,, \quad
\widetilde{(q^{(0,1)A})} = \epsilon_{AB}q^{(1,0)B}\,.\lb{realityq}
\ee

Both sets of constraints are solved in the same way. Consider, for instance,  Eqs. \p{cq1}.
The analyticity conditions (\ref{cq1}a) imply
\be
\mbox{(\ref{cq1}a)} \quad \Rightarrow \quad q^{(1,0)\underline{A}} = q^{(1,0)\underline{A}}(\zeta_+, u, v)\,.
\ee
Then, taking into account the reality conditions \p{realityq}, harmonic constraints (\ref{cq1}b)
leave in $q^{(1,0)\underline{A}}$ just (4 + 4) independent real fields
\be
q^{(1,0)\underline{A}} (\zeta_+, u, v) = f^{i\underline{A}}(t_+)u^1_i + \theta^{1,-1}\psi^{a\underline{A}}(t_+)v^1_a
-\theta^{1,1}\psi^{a\underline{A}}(t_+)v^{-1}_a
-2i\theta^{1,1}\theta^{1,-1} \partial_{t_+}f^{i\underline{A}}u^{-1}_i\,,\lb{q1}
\ee
where we have used the explicit form \p{DHanalB} of $D^{2,0}$ and $D^{0,2}$ in analytic basis \p{A+}.

Quite analogously, in the alternative set \p{cq2}, conditions (\ref{cq2}a) imply the analyticity of the second type  for
$q^{(0,1)A}$,
\be
\mbox{(\ref{cq2}a)} \quad \Rightarrow \quad q^{(0,1)A} = q^{(0,1)A}(\zeta_-, u, v)\,,
\ee
while (\ref{cq2}b) fixes this analytic superfield to have the special form
\be
q^{(0,1)A} (\zeta_-, u, v) = f^{a A}(t_-)v^1_a + \theta^{-1,1}\omega^{i A}(t_-)u^1_i -\theta^{1,1}\omega^{i A}(t_-)u^{-1}_i
-2i\theta^{1,1}\theta^{-1,1} \partial_{t_-}f^{a A}v^{-1}_a\,. \lb{q2}
\ee

Comparing \p{q1} with \p{q2}, we observe that the relevant
irreducible (4 + 4) field sets coincide modulo a permutation of the SU(2)
doublet indices $i \leftrightarrow a$ and $\underline{A}
\leftrightarrow A\,$.

The free superfield actions of these supermultiplets are written as the following integrals over the full bi-HSS:
\be
S_{free}^q \propto \int \mu \left(q^{(1,0)\underline{A}}q^{(-1,0)}_{\underline{A}} - q^{(0,1)A}q^{(0,-1)}_A\right),\lb{qqfree}
\ee
where
\be
q^{(-1,0)\underline{A}} := D^{-2,0}q^{(1,0)\underline{A}}\,, \quad q^{(0,-1)A} := D^{0,-2}q^{(0,1)A}\,.\lb{Defq-}
\ee
Notice that the superfields defined in \p{Defq-} satisfy the relations
\be
D^{2,0}q^{(-1,0)\underline{A}} = q^{(1,0)\underline{A}}\,,\;\; D^{0,2}q^{(0,-1)A} = q^{(0,1)A}\,, \;\;
D^{-2,0}q^{(-1,0)\underline{A}}= D^{0,-2}q^{(0,-1)A} = 0\,,\lb{Constrq-}
\ee
which can be proved using commutation relations \p{HarmComm}, constraints (\ref{cq1}b), (\ref{cq2}b),
and the general relations \p{Lemma1}, \p{Lemma2}.
To avoid possible confusion, let us point out that the two terms in the sum \p{qqfree} are completely independent.
They have been
written together for convenience. The general action of these two multiplets in the bi-HSS is given
by the expression
\be
S^{q}_{gen} \propto \int \mu\, {\cal L}(q^{(\pm 1,0)\underline{A}}, q^{(0, \pm 1)A}, u, v)\,.\lb{qqGen}
\ee
The component form of this general action in the ordinary ${\cal N}{=}4, d{=}1$ superspace has been given in \cite{ILS}.
Its bosonic sector is a sum of the $d{=}1$ pullbacks of a conformally flat metrics for the fields $f^{i\underline{A}}$ and $f^{a A}$ with
two conformal factors related to the superfield Lagrangian. This component action can be easily recovered
from the bi-HSS action \p{qqGen}, by performing $u$ and $v$ harmonic integrations at the final step.

It is interesting that the bi-HSS approach naturally suggests two other forms of the multiplet ${\bf (4,4,0)}$,
such that the extra SU(2) groups, realized on indices $\underline{A}$ and $A$ and commuting with supersymmetry, are
replaced by the complementary SU(2) automorphism groups. The physical bosonic fields, in this case, transform
as a four-vector of the full automorphism group ${\rm SO}(4) \sim {\rm SU}(2)_L\times {\rm SU}(2)_R$
of the ${\cal N}{=}4, d{=}1$ superalgebra.
Both forms of this supermultiplet  are represented by biharmonic superfield $q^{(1,1)}$
which is subjected either to the analyticity conditions (\ref{cq1}a) or to the analyticity conditions (\ref{cq2}a) and to the same
set of harmonic constraints:
\bea
&& \mbox{(a)} \;D^{1,1} q^{(1,1)}_I = D^{1,-1}q^{(1,1)}_I = 0\,, \quad
\mbox{(b)} \; D^{2,0}q^{(1,1)}_I =  D^{0, 2}q^{(1,1)}_I = 0\,, \lb{cq4} \\
&& \mbox{(a)} \;D^{1,1} q^{(1,1)}_{II} = D^{-1,1}q^{(1,1)}_{II} = 0\,, \quad
\mbox{(b)} \; D^{2,0}q^{(1,1)}_{II} =  D^{0, 2}q^{(1,1)}_{II} = 0\,. \lb{cq5}
\eea
The solutions of these constraints can be expressed as
\bea
q^{(1,1)}_I &=& f^{ia}u^1_i v^1_a + \theta^{1,1}\left(\psi + \psi^{(ab)}v^1_av^{-1}_b\right) - \theta^{1,-1}\psi^{(ab)} v^1_a v^1_b
-2i \theta^{1,1}\theta^{1,-1}\partial_{t_+} f^{ia}u^{-1}_iv^1_a\,, \lb{qI}\\
q^{(1,1)}_{II} &=& \hat{f}{}^{ia}u^1_i v^1_a + \theta^{1,1}\left(\omega + \omega^{(ik)}u^1_iu^{-1}_k\right)
- \theta^{-1,1}\omega^{(ik)} u^1_i u^1_k
-2i \theta^{1,1}\theta^{-1,1}\partial_{t_-} \hat{f}{}^{ia}u^{1}_iv^{-1}_a \lb{qII}
\eea
(correspondingly, in the bases \p{A+} and \p{A-}).

We observe that these superfields can indeed be recovered by identifying both the doublet index $\underline{A}$
in \p{q1} with $a\,$, and $A$ in \p{q2}
with $i$. More precisely
\be
q_I^{(1,1)} = q^{(1,0)a}v^1_a\,, \quad q_{II}^{(1,1)} =  q^{(0,1)i}u^1_i\,.
\ee
The free actions are given by the formula
\be
\tilde{S}^q_{free} \propto \int \mu \left(q^{(1,1)}_Iq^{(-1,-1)}_I  - q^{(1,1)}_{II}q^{(-1,-1)}_{II}\right), \lb{qqfree1}
\ee
where
\be
q^{(-1,-1)}_{I, II} := D^{-2,0}D^{0,-2}q^{(1,1)}_{I, II}\,.
\ee
The general action of these superfields is a particular case of \p{qqGen} corresponding to the above-mentioned identification
of the SU(2) groups.

A few further comments are needed.
\begin{itemize}
\item
By making use of the general relations \p{Lemma1}, \p{Lemma2}, the harmonic constraints (\ref{cq1}b) and (\ref{cq2}b)
imply
\be
\mbox{(a)} \;\;D^{0,-2}q^{(1,0)\underline{A}} = 0 \quad \mbox{and} \quad \mbox{(b)} \;\;D^{-2,0}q^{(0,1)A} = 0\,,
\quad \mbox{respectively}.\lb{Bas2}
\ee
\item By the same reasoning, the second analyticity conditions in (\ref{cq1}a) and (\ref{cq2}a)
follow from the first analyticity conditions,
\be
D^{1,1}q^{(1,0)\underline{A}} = D^{1,1}q^{(0,1)A} = 0\,, \lb{Basic}
\ee
and the harmonic constraints (including \p{Bas2}).
\item
A nonlinear generalization of multiplet ${\bf (4,4,0)}$ proposed in \cite{DI,DInew} amounts to the following set of constraints
in the bi-HSS:
\bea
\mbox{(a)} \;D^{1,1} q^{(1,0)\underline{A}} = 0\,, \quad
\mbox{(b)} \; D^{2,0}q^{(1,0)\underline{A}} =  {\cal F}^{(3,0)\underline{A}}(q^{(1,0)}, u, v)\,,
\;D^{0, 2}q^{(1,0)\underline{A}} = 0 \lb{cq3}
\eea
(and to the analogous one for $q^{(0,1)A}\,$). The second harmonic constraint in (\ref{cq3}b) also implies
$D^{0, -2}q^{(1,0)\underline{A}} = 0\,$. From these two constraints it follows that ${\cal F}^{(3,0)\underline{A}}$ does not
involve an explicit dependence on harmonics $v^{\pm 1}_a$, namely,
$\partial^{0\pm 2}{\cal F}^{(3,0)\underline{A}}= 0 \;\Rightarrow \; {\cal F}^{(3,0)\underline{A}} =
{\cal F}^{(3,0)\underline{A}}(q^{(1,0)}, u)\,.$
\item The supermultiplets carried by  $q^{(0,1)A}\,$, $q^{(1,0)\underline{A}}$ provide an ${\cal N}{=}4$ superfield realization
of the ${\cal N}{=}8$ multiplet
${\bf (8,8,0)}\,$\cite{BIKL,ILS}. The second hidden ${\cal N}{=}4$ supersymmetry completing the manifest
${\cal N}{=}4, d{=}1$ supersymmetry
to ${\cal N}{=}8, d{=}1$ is realized as the following transformations of this superfield pair:
\be
\delta q^{(0,1)A} = \varepsilon^A_{\;\;\underline{A}}\,D^{-1,1}q^{(1,0)\underline{A}}\,, \quad \delta q^{(1,0)\underline{A}} =
-\varepsilon_A^{\;\;\underline{A}}D^{1,-1}q^{(0,1)A}\,,\lb{HidN4}
\ee
where $\varepsilon^{A\underline{A}}$ is the corresponding Grassmann parameter. It is easy to check that these transformations
are compatible with the constraints \p{cq1}, \p{cq2}  and that the action \p{qqfree}
is invariant modulo a total harmonic derivative in the integrand. General
conditions of the ${\cal N}{=}8$ invariance of the general action \p{qqGen} (in a formulation through the ordinary
${\cal N}{=}4, d{=}1$ superfields)
were derived in \cite{ILS}. Notice that the transformation laws \p{HidN4} together with those of the manifest ${\cal N}{=}4$
supersymmetry are covariant with respect to the hidden $\rm{SO}(8)/[\rm{SO}(4)\times \rm{SO}(4)]$ transformations \cite{ILS},
in accordance with the property that
the full automorphism group of the ${\cal N}{=}8, d{=}1$ supersymmetry is $\rm{SO}(8)\,$.
\item
Similarly, a hidden ${\cal N}{=}4$ supersymmetry can be realized in terms of superfields $q^{(1,1)}_{I,II}$ (under the
above mentioned identifications of the SU(2) groups)
\bea
\delta q^{(1,1)}_{II} &=& \hat{\varepsilon}{}^{1,-1}D^{-1,1}q^{(1,1)}_I - \hat{\varepsilon}{}^{1,1}D^{-1,1}D^{0,-2}q^{(1,1)}_I\,, \nn
\delta q^{(1,1)}_{I} &=& -\hat{\varepsilon}{}^{-1,1}D^{1,-1}q^{(1,1)}_{II} + \hat{\varepsilon}{}^{1,1}D^{1,-1}D^{-2,0}q^{(1,1)}_{II}\,,
\eea
where $\hat{\varepsilon}{}^{1,1} = \hat{\varepsilon}{}^{ia}u^1_iv^1_a,\; \hat{\varepsilon}{}^{1,-1}
= \hat{\varepsilon}{}^{ia}u^1_iv^{-1}_a\,,$ etc. One can easily check
that these transformations are perfectly compatible with the constraints \p{cq4}, \p{cq5}.
\end{itemize}

Finally, let us dwell on the superconformal properties of the above ${\bf (4,4,0)}$ supermultiplets.

On the superfields  $q^{(1,0)\underline{A}}, q^{(0,1)A}$, like on $q^{(1,1)}_{I,II}$, one can realize
the ${\cal N}{=}4$ superconformal group $D(2,1;\alpha)$ considered in the previous section. Let us, for example,
consider $q^{(1,0)\underline{A}}\,.$
Based upon the coordinate transformation laws \p{TranCooan2}, \p{TranHarm2} with superparameters \p{Aa} - \p{Ae},
it can be checked that analyticity conditions (\ref{cq1}a)
together with harmonic constraints (\ref{cq1}b) are in fact covariant under the whole infinite-dimensional large superconformal group,
provided that $q^{(1,0)\underline{A}}$ is transformed as
\be
\delta q^{(1,0)\underline{A}} \simeq q^{(1,0)\underline{A}}{\,}' (\zeta',u',v') - q^{(1,0)\underline{A}}(\zeta,u,v)
= \tilde{\Lambda}_L q^{(1,0)\underline{A}}\,. \lb{qTranI}
\ee
One can also show that, in the considered case, realizations \p{aa} - \p{af} and \p{Aa} - \p{Ae} are equivalent modulo
harmonic constraint (\ref{Bas2}a) which is a consequence of (\ref{cq1}b). Indeed, as was already mentioned,
the basic constraints for $q^{(1,0)\underline{A}}$ are
\p{Basic} and (\ref{cq1}b). Assuming that $q^{(1,0)\underline{A}}$ transforms with a nonzero weight,
\be
\delta q^{(1,0)\underline{A}} = \Lambda_L\, q^{(1,0)\underline{A}}\,, \lb{qTranII}
\ee
these constraints are manifestly covariant under the three-theta analyticity-preserving variations with the
superparameters \p{aa} - \p{af}.
Then, taking the active interpretation of the same full variation of  $q^{(1,0)\underline{A}}$, one finds that it differs from
the variation corresponding to the two-theta
analytic superparameters \p{Aa} - \p{Ae} merely by terms proportional to $D^{0,-2}q^{(1,0)\underline{A}}$ which are zero by virtue
of (\ref{Bas2}a).
Hence, in the realization on $q^{(1,0)\underline{A}}$, one can identify the superparameters in both realizations,
in particular, $\tilde{\Lambda}_L$ with $\Lambda_L$. In the same way, due to the constraint $D^{-2,0}q^{(0,1)A} =0\,$,
the supergroup \p{aa} - \p{af}, in the realization on $q^{(0,1)A}\,$,
can be identified with a realization which is mirror to \p{Aa} - \p{Ae} and preserves the alternative
two-theta analytic subspace  \p{Asubs-}.
The transformation law of $q^{(0,1)A}$ is
\be
\delta q^{(0,1) A} = \tilde{\Lambda}_R\, q^{(0,1) A} = \Lambda_R\, q^{(0,1) A}\,.\lb{qTranIII}
\ee
Notice that $\Lambda_R$ ``lives'' on the second two-theta analytic superspace \p{Asubs-}, so that \p{qTranIII} is compatible with
the constraints \p{cq2}.

The superfield $q^{(1,1)}_I$ and its mirror counterpart $q^{(1,1)}_{II}$ do not satisfy any extra harmonic constraints of the type \p{Bas2},
and, for this reason, one can implement on them only supergroup \p{Aa} - \p{Ae} and its mirror. For instance, in the case of $q^{(1,1)}_I\,$,
\be
\delta q^{(1,1)}_I = \tilde{\Lambda}_L\, q^{(1,1)}_I\,.
\ee
Constraints \p{cq4} are covariant with respect to these transformations. They are covariant also under the global SU(2)
with parameters $\hat{\tau}^{(ab)}$ realized on harmonics $v^{\pm 1}_a$.
The corresponding weight transformation of $q^{(1,1)}_I$ is
\be
\hat{\delta} q^{(1,1)}_I = (\hat{\tau}^{(ab)}v^1_av^{-1}_b)\, q^{(1,1)}_I\,.
\ee

While defining constraints of $q$ superfields are covariant under the infinite-dimensional large
${\cal N}{=}4$ superconformal group, the $q$-superfield actions can be invariant only under the finite-dimensional
${\cal N}{=}4$ superconformal group $D(2,1;\alpha)\,$ - for some special values of parameter $\alpha\,$. For instance,
using transformation laws \p{qTranII}, \p{qTranIII}, \p{TranMeI}, \p{D-I} and \p{D-II}, as well as the Grassmann
analyticity conditions together with the relations \p{Constrq-} and \p{Bas2},
one can directly check that the $D(2,1;\alpha)$ variations of two separate terms in the free action \p{qqfree}
can be made vanishing (up to a total harmonic derivative), but for different
choices of parameter $\alpha\,$. The first term is superconformal only for $\alpha =1$,
while the second one only for $\alpha = -2$.\footnote{This difference in the superconformal properties of $q^{(1,0)\underline{A}}$
and $q^{(0,1)A}$ can be understood from the following simple reasoning. The dilatation weight of integration measure $\mu$
is $-1\,$, while the weights of $q^{(1,0)\underline{A}}$ and $q^{(0,1)A}$ are $\alpha/2$ and $-(1+\alpha)/2\,$, respectively.
This follows from the form of the superfield weight factor $\tilde{\Lambda}_L$ in \p{lambda}
in which substitutions \p{sconfcond} have been made, and its ``mirror'' counterpart $\tilde{\Lambda}_R$ obtained
from $\tilde{\Lambda}_L$ via the changes \p{mnem}.
Then, in order to cancel the dilatation weight of the measure, the weights of both terms in Lagrangian \p{qqfree} should be
+1, and this is achieved with $\alpha = 1$ for the first term and $\alpha = -2$ for the second term.} From this result it follows,
in particular, that the total free action \p{qqfree},
though being invariant under the hidden ${\cal N}{=}8$ supersymmetry, is not ${\cal N}{=}4$ superconformal and so is not ${\cal N}{=}8$
superconformal either.
A similar situation takes place for the free actions of superfields $q^{(1,1)}_I$ and $q^{(1,1)}_{II}$ in \p{qqfree1}: one can show
that the $q^{(1,1)}_I$ action is invariant under realization \p{alphaII} of $D(2,1;\alpha)$ with $\alpha = 1\,$, whereas the
$q^{(1,1)}_{II}$ action is invariant under the mirror image of \p{alphaII} with $\alpha = -2\,$.

In fact, the $D(2,1; \alpha)$ invariant actions of ${\bf (4 , 4, 0)}$ multiplets of both sorts can be constructed for any $\alpha$,
but, in the generic case, these actions involve some sigma-model-type self-interactions \cite{IKLe,IL,ikl1} (see also \cite{Str,Po}).
Thus one can hope
to construct the ${\cal N}{=}8$ superconformal actions by combining those $q^{(1,0)\underline{A}}$ and $q^{(0,1)A}$ actions
which are ${\cal N}{=}4$ superconformal for the same value of $\alpha$. A candidate action of this type
was presented in \cite{ILS} in the ordinary ${\cal N}{=}4$ superfield approach. An interesting problem
for the future is to reformulate it in the bi-HSS and to examine its ${\cal N}{=}4$ and ${\cal N}{=}8$ superconformal properties.

\subsection{Multiplets ${\bf (3,4,1)}$}
Once again, there are two types of biharmonic superfields accommodating the off-shell multiplet ${\bf (3,4,1)}$, viz. $W^{(2,0)}$
and $W^{(0,2)}\,$. They differ in the type of Grassmann analyticity:
\bea
\mbox{(a)} \;D^{1,1} W^{(2,0)} = D^{1,-1} W^{(2,0)} = 0\,, \quad
\mbox{(b)} \; D^{2,0}W^{(2,0)} =  D^{0, 2,}W^{(2,0)} = 0 \lb{cw1}
\eea
and
\bea
\mbox{(a)} \;D^{1,1} W^{(0,2)} = D^{-1,1} W^{(0,2)} = 0\,, \quad
\mbox{(b)} \; D^{2,0}W^{(0,2)} =  D^{0, 2,}W^{(0,2)} = 0\,. \lb{cw2}
\eea
The solutions of \p{cw1} and \p{cw2} read
\bea
W^{(2,0)}(\zeta_+,u,v) &=& w^{(ik)}(t_+)u^1_iu^1_k + \theta^{1,-1}\psi^{ia}(t_+)u^1_iv^1_a -\theta^{1,1}\psi^{ia}(t_+)u^1_iv^{-1}_a \nn
&& +\, \theta^{1,1}\theta^{1,-1}\left[w(t_+) -2i\partial_{t_+}w^{(ik)}u^1_iu^{-1}_k\right], \lb{cw3} \\
W^{(0,2)}(\zeta_-,u,v) &=& \hat{w}{}^{(ab)}(t_-)v^1_av^1_b + \theta^{-1,1}\hat{\psi}{\,}^{ia}(t_-)u^1_iv^1_a
-\theta^{1,1}\hat{\psi}{\,}^{ia}(t_-)u^{-1}_iv^{1}_a \nn
&& +\, \theta^{1,1}\theta^{-1,1}\left[\hat{w}(t_-) -2i\partial_{t_-}\hat{w}{}^{(ab)}v^1_av^{-1}_b\right]. \lb{cw4}
\eea
Thus, the irreducible ${\bf (3, 4, 1)}$ field sets are $(w^{(ik)}, \psi^{ia}, w)$ and $(\hat{w}{}^{(ab)},
\hat{\psi}{\,}^{ia}, \hat{w})$, and they differ merely in the SU(2) assignment of physical bosonic fields which
form triplets of the automorphism groups ${\rm SU}(2)_L$ and ${\rm SU}(2)_R\,$, respectively.

In order to construct the invariant actions, one should define the full set of nonanalytic harmonic projections
of the basic analytic superfields:
\be
W = D^{-2,0}W^{(2,0)}\,, \; W^{(-2,0)} = (D^{-2,0})^2W^{(2,0)}\,, \; \hat{W} = D^{0,-2}W^{(0,2)}\,, \;
W^{(0,-2)} = (D^{0,-2})^2W^{(0, 2)}\,.
\ee
Notice that
\be
D^{0, -2}W^{(2,0)} = D^{-2, 0}W^{(0,2)} =0\,, \;\; D^{-2,0} W^{(-2,0)} = D^{0,-2}W^{(0,-2)} = 0
\ee
as a consequence of the harmonic constraints (\ref{cw1}b), (\ref{cw2}b). The free action and the most general sigma-model-type action
of this superfield system are
\bea
&& S^W_{free} \propto \int \mu \left(W^{(2,0)}W^{(-2,0)} - W^{(0,2)}W^{(0, -2)}\right), \lb{Wfree} \\
&& S^W_{gen} \propto \int \mu\,{\cal L}(W^{(\pm 2, 0)}, W^{(0,\pm 2)}, W, \hat{W})\,. \lb{Wgen}
\eea
The component structure of these actions in the ordinary ${\cal N}{=}4$ superspace was presented in \cite{ILS}. The actions
in the bi-HSS give rise to the same component actions (performing the $u$ and $v$ harmonic integrals at the final steps).

On these two ${\cal N}{=}4$ multiplets, one can also implement the ${\cal N}{=}8$ supersymmetry with respect to which they
combine into an off-shell ${\bf (6, 8, 2)}$ multiplet \cite{ILS}. The transformations of the second hidden ${\cal N}{=}4$
supersymmetry on the biharmonic superfields $W^{(2,0)}, W^{(0,2)}$ are
\bea
&& \delta W^{(2,0)} = \hat{\varepsilon}{}^{1,-1}D^{1,-1}W^{(0,2)} - \hat{\varepsilon}{}^{1,1}D^{1,-1}D^{0,-2} W^{(0,2)}\,, \nn
&& \delta W^{(2,0)} = -\hat{\varepsilon}{}^{-1,1}D^{-1,1}W^{(2,0)} + \hat{\varepsilon}{}^{1,1}D^{-1,1}D^{-2,0} W^{(2,0)}\,. \lb{HidN42}
\eea
These transformations are compatible with the $W^{(2,0)}, W^{(0,2)}$ defining constraints \p{cw1}, \p{cw2}. The free action \p{Wfree}
is invariant with respect to them up
to a total derivative in the integrand. Conditions of the ${\cal N}{=}8$ invariance of a general sigma-model-type action \p{Wgen}
were derived in \cite{ILS} in the framework of the conventional ${\cal N}{=}4$ superfield description.

Constraints \p{cw1} and \p{cw2} are covariant under the infinite-dimensional large ${\cal N}{=}4$ $\rm{SO}(4)\times \rm{U}(1)$ superconformal
group in the realization \p{Aa} - \p{Ae} and in its mirror counterpart, respectively, provided that the superfields
$W^{(2,0)}$ and $W^{(0,2)}$ transform as
\be
\delta W^{(2,0)} = 2\tilde{\Lambda}_L \,W^{(2,0)}\,, \quad \delta W^{(0,2)} = 2\tilde{\Lambda}_R\, W^{(0,2)}\,.
\ee
Like in the case of multiplets ${\bf (4,4,0)}$, the $W^{(2,0)}$ and $W^{(0,2)}$ superfield actions can be invariant
only under the finite-dimensional ${\cal N}{=}4$ superconformal symmetry $D(2,1;\alpha)$. The free action of $W^{(2,0)}$ in \p{Wfree}
is invariant with respect to $D(2,1;\alpha = \frac{1}{2})$, while that of $W^{(0,2)}$ with respect to $D(2,1;\alpha = -\frac{3}{2})$.
This can be shown by exploiting the superconformal transformations with the superparameters \p{alphaII} and their mirror counterparts, and
by making use of the Grassmann analyticity constraints in \p{cw1}, \p{cw2}, as well as the harmonic constraints together
with some of their consequences, e.g.,
\be
\mbox{(a)} \;\;D^{0,-2}\,W^{(2,0)} = D^{-2,0}\,W^{(0, 2)} = 0\,, \quad \mbox{(b)} \;\;(D^{-2, 0})^3 \,W^{(2,0)}
= (D^{0, -2})^3 \,W^{(0,2)} = 0\,. \lb{AddWconstr}
\ee
The superconformal actions, for any other choice of $\alpha$, can also be constructed, but they involve necessarily
interactions \cite{IKL0,IL,ikl1}.
Let us remark that, as a consequence of the constraints (\ref{AddWconstr}a), the realizations \p{alphaI} and \p{alphaII}
(for the relevant choices of $\alpha$) are equivalent to each other when applied to $W^{(2,0)}$ and $W^{(0,2)}$
(as in the case of $q^{(1,0)\underline{A}}$ and $q^{(0,1)A}$\,).

\subsection{Multiplets ${\bf (1,4,3)}$}
The multiplet ${\bf (1,4,3)}$ and its mirror are defined by constraints of the second order in the spinor
derivatives \cite{IKLe,IKP}. In the bi-HSS these multiplets are described by zero-charge superfields ${\cal U}$ and ${\cal V}$
defined by the following constraints:
\bea
&& \mbox{(a)} \;\; D^{1,1}D^{1,-1} {\cal U} = c{\,}^{2, 0}\,, \quad \mbox{(b)}\;\; D^{2,0}{\cal U} = D^{0,2}{\cal U} = 0\,, \lb{u} \\
&& \mbox{(a)} \;\; D^{1,1}D^{-1,1} {\cal V} = \tilde{c}{\,}^{0,2}\,, \quad \mbox{(b)}\;\; D^{2,0}{\cal V}
= D^{0,2}{\cal V} = 0\,, \lb{tildeu}
\eea
where
\be
c{\,}^{2,0} = c^{(ik)}u^1_iu^1_k\,, \quad \tilde{c}{\,}^{0,2} = \tilde{c}{\,}^{(ab)}v^1_av^1_b\,.
\ee
Here, $c^{(ik)}, \tilde{c}{}^{(ab)}$ are two independent constant triplets which break the ${\rm SU}(2)_L$
and/or ${\rm SU}(2)_R$ symmetries.
It is self-consistent to choose both or one of these triplets equal to zero.

Solution of the constraints \p{u} can be written in basis \p{A+} in the form:
\bea
{\cal U} = U(\zeta_+,u,v) + \theta^{-1,-1}\Omega^{1,1}(\zeta_+, u, v) - \theta^{-1,1}\Omega^{1,-1}(\zeta_+, u, v)
+  \theta^{-1,-1}\theta^{-1,1} c^{(ik)}u^1_iu^1_k\,,    \lb{u1}
\eea
where
\bea
U  &=& \phi(t_+) + \theta^{1,1}\chi^{ia}(t_+)u^{-1}_iv^{-1}_a - \theta^{1,-1}\chi^{ia}(t_+)u^{-1}_iv^{1}_a
-  \theta^{1,1}\theta^{1,-1} c^{(ik)}u^{-1}_iu^{-1}_k\,,    \lb{U} \\
\Omega^{1,1} &=& \chi^{ia}u^{1}_iv^{1}_a + \theta^{1,-1}\phi^{(ab)}v^1_a v^1_b + \theta^{1,1}\left(\partial_{t_+}\phi
- c^{(ik)}u^1_iu^{-1}_k -
\phi^{(ab)}v^{-1}_a v^1_b \right)\nn
&& -\,2i \theta^{1,1}\theta^{1,-1} \partial_{t_+}\chi^{ia}u^{-1}_iv^1_a\,, \nn
\Omega^{1,-1} &=& \chi^{ia}u^{1}_iv^{-1}_a - \theta^{1,1}\phi^{(ab)}v^{-1}_a v^{-1}_b
+ \theta^{1,-1}\left(\partial_{t_+}\phi - c^{(ik)}u^1_iu^{-1}_k +
\phi^{(ab)}v^{-1}_a v^1_b \right)\nn
&& -\,2i \theta^{1,1}\theta^{1,-1} \partial_{t_+}\chi^{ia}u^{-1}_iv^{-1}_a\,. \lb{Om}
\eea
Thus, the irreducible ${\bf (1, 4, 3)}$ field content of the multiplet in question is the set of (4 + 4) fields
$(\phi(t), \phi^{(ab)}(t), \chi^{ia}(t))$, as it should be.

Solution ${\cal V}$ of the constraints \p{tildeu} is naturally written
in basis \p{A-}. It is obtained from \p{u1} - \p{Om} by the formal changes (cf. \p{mnem})
\be
t_+\;\rightarrow \;t_-\,, \quad  i \;\leftrightarrow \; a\,, \quad u^{\pm 1}_i \;\leftrightarrow \; v^{\pm 1}_a\,, \quad
\theta^{1, -1} \;\leftrightarrow \; \theta^{-1, 1}\,.
\ee
The corresponding irreducible field content is $(\tilde{\phi}(t), \tilde{\phi}^{(ik)}(t), \tilde{\chi}^{ia}(t))$. Thus
these two off-shell ${\bf (1, 4, 3)}$ multiplets differ in the SU(2) assignment of auxiliary bosonic
fields which form triplets of either ${\rm SU}(2)_L$ or ${\rm SU}(2)_R$.

The free and general actions of these two multiplets are given by the following integrals over the bi-HSS
\bea
&& S_{free} \propto \int \mu \left({\cal U}^2 - {\cal V}^2\right), \nn
&& S_{gen} \propto \int \mu\, {\cal L}({\cal U}, {\cal V}, u, v)\,. \lb{Suv}
\eea
On this pair of biharmonic superfields one can also realize a hidden second ${\cal N}{=}4$ supersymmetry which
extends the manifest ${\cal N}{=}4$ supersymmetry to ${\cal N}{=}8$. The $D(2,1; \alpha)$ superconformal properties
of these multiplets were also studied in detail
(see e.g. \cite{IKLe,ikl1,DI1}), and they can be easily translated into the bi-HSS language.

\subsection{An example of a new off-shell ${\cal N}{=}4$ supermultiplet}
The off-shell ${\cal N}{=}4$ multiplets discussed above are not new; they admit an alternative description either in
the ordinary ${\cal N}{=}4$ superspace
\cite{IKLe,ikl1,bk}, or in the harmonic ${\cal N}{=}4$ superspace with one set of harmonic variables \cite{IL,DI,DI1}.
An advantage of the bi-HSS is that it gives the possibility of the joint description of these multiplets together
with their mirror counterparts. The basic new feature of the bi-HSS
is the presence of a new analytic subspace in it -- the three-theta analytic superspace \p{Asubs3}.
This property provides an opportunity
to define new off-shell ${\cal N}{=}4$ superfields. Namely, one can define three-theta
analytic superfields $G^{(p, q)}(\zeta, u, v), \;p> 0, q>0\,,$
\be
\mbox{(a)} \; D^{1,1}G^{(p, q)} = 0; \qquad \mbox{(b)} \; D^{2,0}G^{(p, q)} = D^{0,2}G^{(p, q)} = 0\,.\lb{ConstrNew}
\ee
If $q$ or $p$ is zero, \p{ConstrNew} would imply $D^{1,-1}G^{(p,0)} = 0$ or $D^{-1,1}G^{(0,q)} = 0\,$; i.e.
the corresponding superfields would be automatically two-theta analytic and would be reduced either to multiplets
$q^{(1,0)}, W^{(2,0)}, q^{(0,1)}, W^{(0,2)}$ treated in the previous subsections  or
to some direct generalizations of the latter with $p>2, \,q=0$ or $p = 0, \,q >2$ (considered in \cite{IL} in the
framework of the standard ${\cal N}{=}4, d{=}1$ HSS). However, for $p,\,q \neq 0\,$, the second Grassmann harmonic analyticity conditions
do not directly follow from \p{ConstrNew}. So, in this case these constraints define some new off-shell representations
of the ${\cal N}{=}4, \,d{=}1$ supersymmetry.

Here, we consider the simplest example of such a superfield, namely, that corresponding to the choice $p = q =1$. Constraints \p{ConstrNew}
uniquely fix the component field content of $G^{(1,1)}$:
\bea
G^{(1,1)}(\zeta, u, v) &=& f^{ia}u^1_iv^1_a + \theta^{1,1}\left[\psi + \psi^{(ab)}v^1_av^{-1}_b +  \psi^{(ik)}u^1_iu^{-1}_k\right]
- \theta^{1,-1}\psi^{(ab)}v^1_av^{1}_b \nn
&& -\, \theta^{-1,1}\psi^{(ik)}u^1_iu^{1}_k  + i\theta^{1,1}\theta^{1,-1}\,(g^{ia}- \dot{f}{}^{ia})u^{-1}_iv^{1}_a
- i\theta^{1,1}\theta^{-1,1}\,(g^{ia} + \dot{f}^{ia})u^{1}_iv^{-1}_a \nn
&& -\, i\theta^{-1,1}\theta^{1,-1}\,(g^{ia} + \dot{f}^{ia})u^{1}_iv^{1}_a
+ i\theta^{1,1}\theta^{1,-1}\theta^{-1,1}\, [\omega + \dot{\psi} +2\dot{\psi}^{(ik)}u^1_iu^{-1}_k ]\,. \lb{SolG}
\eea
Thus we have an $(8 + 8)$ off-shell representation consisting of eight bosonic $d{=}1$ fields  $f^{ia}(t), g^{ia}(t)$ and eight
fermionic fields $\psi(t), \psi^{(ab)}(t), \psi^{(ik)}(t), \omega(t)\,$. By dimensionality reasoning,
the fields $f^{ia}(t)$, $\psi(t), \psi^{(ik)}(t)$ and $\psi^{(ab)}(t)$ are candidates for physical fields, and
$g^{ia}(t)$ and $\omega(t)$  for auxiliary fields. Note that $\psi^{(ab)}$ and $\psi^{(ik)}$ appear in \p{SolG}
in an asymmetric way, though one would expect them to be on equal footing. This asymmetry is in fact an artifact
of our choice of the $t_+$ basis \p{A+} in the three-theta superspace; after passing to the basis \p{A-}, the $\theta$ expansion in \p{SolG}
takes the form in which $\psi^{(ab)}$ and $\psi^{(ik)}$ exchange their places. The ${\cal N}{=}4, d{=}1$
supersymmetry is realized on these $(8 +8)$ fields as follows:
\bea
&& \delta f^{ia} = -\varepsilon^{ia}\,\psi + \varepsilon^i_{\;\;b}\,\psi^{(ab)} + \varepsilon^{\;\;a}_k\,\psi^{(ik)}\,, \quad
\delta g^{ia} = -\varepsilon^{ia}\,\omega - \varepsilon^i_{\;\;b}\,\dot{\psi}{}^{(ab)} + \varepsilon^{\;\;a}_k\,\dot{\psi}{}^{(ik)}\,, \nn
&& \delta \psi^{(ik)} = i \varepsilon^{(i a} \,( g{\,}^{k)}_{\;a} + \dot{f}{\,}^{k)}_{\;a})\,, \quad
\delta \psi^{(ab)} = -i \varepsilon^{i (a} \,(g_{i}{\,}^{\;b)} - \dot{f}_{i}{\,}^{\;b)}), \quad \delta \psi =
-i \varepsilon^{ia}\dot{f}{}_{ia}\,, \nn
&& \delta \omega = -i \varepsilon^{ia}\dot{g}{}_{ia}\,. \lb{TranGc}
\eea

Without loss of generality, the free action $S_{G}^{free}$ can be chosen in the form
\be
S_{G}^{free} \propto \int \mu \, G^{(-1,1)} G^{(1,-1)}\,, \;\mbox{where}\;  G^{(-1,1)}: = D^{-2,0}G^{(1,1)}\,, \;\;
G^{(1,-1)}: = D^{0,-2}G^{(1,1)}\,. \lb{LagrG}
\ee
Alternative bilinear superfield Lagrangians, e.g., $G^{(1,1)}G^{(-1,-1)} = G^{(1,1)} D^{-2,0}D^{0,-2}G^{(1,1)}\,$, are either reduced
to \p{LagrG} via integration by parts, or are vanishing as a consequence of the constraints \p{ConstrNew}
and their corollaries
$$
(D^{-2,0})^2 G^{(1,1)} = (D^{0,-2})^2 G^{(1,1)} = 0\,.
$$
After substituting the precise form of $G^{(-1,1)}$ and $G^{(1,-1)}$ into action \p{LagrG} and integrating over
the $\theta$ variables and harmonics $u$ and $v$, one gets the following component off-shell form of \p{LagrG}:
\be
S_G^{free} \propto \int dt \left(i \psi^{(ik)}\dot{\psi}{}_{(ik)} - i \psi^{(ab)}\dot{\psi}{}_{(ab)} -2i\omega \psi
+ 2 g^{ia}\dot{f}_{ia}\right). \lb{LagrGcomp}
\ee
We see that $\omega$ and $\psi$ form a pair of auxiliary fermionic fields with complementary dimensions;
the only physical fermionic fields are $\psi^{(ik)}$ and $\psi^{(ab)}$. Surprisingly, in the bosonic sector,
instead of the standard kinetic term for field $f^{ia}$, i.e. $\dot{f}^{ia}\dot{f}_{ia}\,$, we find a Lagrangian
of the first order in the time derivative, giving rise
to the first-order equations of motion:
$$
\dot{g}^{ia} = \dot{f}^{ia} = 0\,.
$$
It can be interpreted as a sort of the $d{=}1$ Wess-Zumino (or Chern-Simons) Lagrangian describing a Lorentz coupling of the target coordinate
$f^{ia}(t)$ to some external ``magnetic'' potential $g^{ia}(t)$. We can rescale $g^{ia}$ as $g^{ia}= 2\kappa \tilde{f}^{ia}$,
$[\kappa] = cm^{-1}$, and, consequently, pass to the doubled  eight-dimensional target coordinate set
$(\tilde{f}^{ia}, f^{ia}) := f^{ia}_\mu\,, \;\mu=1,2$.
After this redefinition, the WZ term in \p{LagrGcomp} (modulo a total time derivative) takes the following more familiar form:
\be
g^{ia}\dot{f}_{ia} = \kappa \epsilon^{\mu\nu}f^{ia}_\mu \dot{f}_{ia\,\nu}\,, \quad \epsilon^{\mu\nu} = -\epsilon^{\nu\mu}, \;\epsilon^{21} =1\,,
\ee
with $\kappa$ being a constant external magnetic field. The models of such WZ (or CS \cite{HT}) mechanics received some attention
in connection with the famous Landau problem (see, e.g., \cite{Land}, \cite{Ivan} and references therein), as well as with the matrix models
(see, e.g., \cite{Poly}). Thus, the off-shell $(8+8)$ multiplet \p{SolG} living on the three-theta ${\cal N}{=}4, d{=}1$ analytic superspace
naturally gives rise to an ${\cal N}{=}4$ superextension of the simple $d{=}1$ WZ mechanics with a specific eight-dimensional target manifold.
Notice the relative minus sign between the fermionic kinetic terms in \p{LagrGcomp}; it signals the presence of fermionic ghost states
in quantum theory, though everything  is self-consistent at the classical level.\footnote{In fact, the quantum theory can be
``cured'' by methods similar to those worked out for situations like in \cite{VP} - \cite{CIMT}. We thank S. Fedoruk and
A. Smilga for a discussion of this point.}

It is straightforward to find how \p{LagrGcomp} generalizes to the interacting case. The general action of superfield $G^{(1,1)}\,$,
giving rise to the fermionic kinetic terms with only one time derivative, can be written as \footnote{One could start with a more general
Lagrangian, ${\cal L}(G^{(1,-1)}, G^{(-1,1)}, G^{(1,1)}, G^{(-1,-1)}, u, v)\,$, which is expandable in series with
respect to its functional arguments, and show that it is indeed reduced, modulo a total harmonic derivative, to \p{LagrGgen}.}
\be
S_G = \int \mu\;{\cal L}(G^{(1,-1)}, G^{(-1,1)}, u, v)\,. \lb{LagrGgen}
\ee
For brevity, we present only the bosonic core of the component off-shell action following from the superfield action \p{LagrGgen}:
\be
S_G^{bos} \sim 2\int dt \,{\cal F}(f)\,g^{ia}\dot{f}_{ia}\,, \quad {\cal F}(f) =
\int du dv \frac{\partial^2 {\cal L}(f^{(1,-1)}, f^{(-1,1)}, u, v)}{\partial f^{(1,-1)}\partial f^{(-1,1)}}\,. \lb{Gencomp}
\ee
We observe that the whole effect of self-interaction of fields $f^{ia}(t)$ can be absorbed into the following redefinition of $g^{ia}\,$:
$$
{\cal F}(f)\,g^{ia} = \tilde{g}^{ia}\,.
$$
Consequently, the above WZ term is form-invariant against passing to a general action of $G^{(1,1)}$.

Reduction to the two-theta analytic supermultiplets $q^{(1,1)}_I$ and $q^{(1,1)}_{II}$ is accomplished by imposing additional
analyticity conditions:
$$
\mbox{(a)}\; D^{1,-1}G^{(1,1)} = 0 \qquad \mbox{or} \quad \mbox{(b)} \;  D^{-1,1}G^{(1,1)} = 0\,,
$$
which yield the following constraints on component fields:
$$
\mbox{(a)}\; \psi^{(ik)} = 0\,, \;g^{ia} = -\dot{f}^{ia}\,, \; \omega = -\dot\psi; \qquad \mbox{(b)} \; \psi^{(ab)} = 0\,, \;
g^{ia} = \dot{f}^{ia}\,, \; \omega = \dot\psi\,.
$$
One can check that they are covariant under the off-shell transformations \p{TranGc}. Action \p{LagrGcomp} is reduced
(up to an overall sign) to the component actions of either $q^{(1,1)}_I$
or $q^{(1,1)}_{II}\,$. The WZ term converts to  the standard kinetic terms of $f^{ia}$. After these reductions, the scalar factor
${\cal F}$ in \p{Gencomp} cannot be removed, in accordance with the fact \cite{IL} that the target geometry of the ${\bf (4,4,0)}$
multiplets is conformally flat.

Constraints \p{ConstrNew}, including the case of $q=p=1$, are covariant under the large ${\cal N}{=}4$ $\rm{SO}(4)\times \rm{U}(1)$
superconformal group in the realization \p{aa} - \p{af}, provided that $G^{(p,q)}$ transforms as
\be
\delta G^{(p, q)} = \left( p\Lambda_L + q\Lambda_R\right) G^{(p, q)}\,.
\ee
One can check that actions \p{LagrG} and \p{LagrGcomp} are not invariant under the supergroup $D(2,1;\alpha)$ for any choice of parameter
$\alpha$, so that they are not superconformal.

\setcounter{equation}{0}
\section{Gauge superfields}
For gauging various isometries of superfield theories in the ${\cal N}{=}4$ bi-HSS, we need appropriate nonpropagating gauge superfields.
For simplicity, we shall consider here the Abelian case. Generalization to the case of non-Abelian isometries is straightforward.

Analogously to the case of the standard ${\cal N}{=}4$ HSS \cite{HSS,book},  we are led to introduce, as basic geometric objects,
the analytic gauge connections
which lengthen the analyticity-preserving  harmonic derivatives. Because there exist three different
types of analytic subspaces in the ${\cal N}{=}4$ bi-HSS, one can define three analytic frames (the so-called $\lambda$ frames)
and, correspondingly, three different types of analytic harmonic gauge connections. As we shall show,
these frames are related to each other and to the $\tau$ frame (in which gauge
parameters are the harmonic-independent ${\cal N}{=}4, d{=}1$ superfields) by appropriate ``bridges.''

\subsection{The first $\lambda$ frame}
First, we consider a frame corresponding to the Grassmann analyticity pattern \p{Asubs+}, \p{Anal+}, \p{CR1}.
In this case, there are three analyticity-preserving harmonic derivatives $(D^{2,0}, D^{0,\pm 2})$. They are covariantized as
\bea
&&\left( D^{2,0}\, \,D^{0, \pm 2}\right) \; \Rightarrow \;  \left( \nabla^{2,0}\, \,\nabla^{0, \pm 2}\right), \quad
\nabla^{2,0} = D^{2,0} - V^{2,0}{\cal J}\,, \; \nabla^{0,\pm 2} = D^{0,\pm 2} - V^{0,\pm 2}{\cal J}\,, \lb{gCon+2} \\
&& D^{1,1}V^{2,0} = D^{1,-1}V^{2,0} = 0\,, \quad D^{1,1}V^{0,\pm 2} = D^{1,-1}V^{0,\pm 2} =0\,.
\eea
Here, ${\cal J}$ is the anti-Hermitian generator of some one-parameter symmetry. The real analytic harmonic
connections $V^{2, 0}$, $V^{0, \pm 2}$ have the following standard gauge transformation laws:
\be
V^{2,0}{\,}' = V^{2,0}  + D^{2,0}\lambda\,, \quad V^{0,\pm 2}{\,}' =  V^{0,\pm 2} + D^{0,\pm 2}\lambda\,, \lb{GaugeV}
\ee
where $\lambda$ is an analytic gauge parameter, $\lambda = \lambda(\zeta_+,u,v)\,$. Requiring the covariantized harmonic
derivatives to satisfy the same algebra as the flat ones, we get a number of harmonic flatness conditions
for analytic connections:
\bea
&& F^{2,2} := D^{2, 0}V^{0, 2} - D^{0, 2}V^{2, 0} = 0\,, \lb{HFa} \\
&& F := D^{0,2}V^{0, -2} -  D^{0, -2}V^{0, 2} = 0\,, \lb{HFb} \\
&& F^{2, -2} := D^{2,0}V^{0, -2} -  D^{0, -2}V^{2, 0} =  0\,. \lb{HFc}
\eea

These additional harmonic constraints for analytic gauge connections are new features of the ${\cal N}{=}4, d{=}1$ bi-HSS
as compared to the ordinary  ${\cal N}{=}4, d{=}1$ HSS \cite{DI}. Let us remark that relation \p{HFc} is a consequence of \p{HFa}
and \p{HFb}. Indeed, using these two conditions and the commutation relations \p{HarmComm}, one can show that
\be
D^{0,2}F^{2, -2} = 0\,.
\ee
Then $F^{2,-2}$ is vanishing due to {\it Proposition} \p{Lemma1}.

Connection $V^{0,-2}$ is not an independent quantity, since it can be expressed through $V^{0,2}$ from \p{HFb}, like the nonanalytic $V^{--}$
connection in the standard HSS is expressed through the analytic one $V^{++}$ \cite{HSS}. The specificity of the bi-HSS is
that $V^{0,-2}$ is analytic like $V^{2,0}$ and $V^{0,2}$. In fact, its analyticity is already implied by analyticity of the
connection $V^{0,2}$, i.e. by the conditions $D^{1,1}V^{0,2} = D^{1,-1}V^{0,2} = 0$. This can be proved using
\p{Lemma1} and relation \p{HFb}.

Thus, the basic quantities of the gauge formalism in the ${\cal N}{=}4$
bi-HSS are the analytic gauge connections $V^{2,0}$ and $V^{0,2}$
subjected to the additional harmonic constraint \p{HFa}. As usual,
to reveal the irreducible field content of these superfields, one
must pass to the WZ gauge. Direct calculations
show that the gauge freedom \p{GaugeV} together
with constraint \p{HFa} allow one to cast both connections, $V^{2,0}$ and
$V^{0,2}\,$, in the following very simple form\footnote{From now on, for brevity, we omit the indices $\pm $ of time variables in
analytic bases.}:
\be
V^{2,0}_{WZ} =
2i\theta^{1,1}\theta^{1,-1} {\cal A}(t)\,, \quad  V^{0,2}_{WZ} = 0\,, \qquad {\cal A}(t){\,}' =
{\cal A}(t) + \dot{\lambda}_0(t)\,,\lb{WZ}
\ee
where $\lambda_0(t)$ is a harmonic-independent part of $\lambda|_{\theta = 0}$. Thus,
we face the same phenomenon as in the case of the
ordinary ${\cal N}{=}4, d{=}1$ HSS \cite{DI}: the only unremovable
component in the off-shell gauge supermultiplet is the bosonic $d{=}1$
``gauge field'' ${\cal A}(t)$ with a residual gauge transformation law as
in \p{WZ}. This transformation law means that {\it locally} the
considered supermultiplet contains $(0 + 0)$ off-shell degrees of
freedom, but ${\cal A}(t)$ cannot be {\it globally} gauged away using the
gauge freedom \p{WZ}. This is the reason why such gauge
supermultiplets were called ``topological'' in \cite{DI}.
The difference of the considered case from the case of
topological gauge multiplets in the ordinary ${\cal N}{=}4, d{=}1$ HSS is that
the underlying analytic biharmonic gauge freedom alone is not sufficient to achieve the
minimal field representation \p{WZ}. In addition, one needs to exploit
the harmonic flatness relation \p{HFa}. Notice that condition \p{HFb}  implies
that  the remaining analytic connection $V^{0,-2}$ is also vanishing in gauge \p{WZ}:
\be
V^{0,-2}_{WZ} = 0\,. \lb{WZ2}
\ee

As in the standard HSS, having at hand the analytic potentials $V^{2,0}\,$, $V^{0,2}\,$,
one can define a nonanalytic connection $V^{-2,0}$ which covariantizes the remaining harmonic derivative
$D^{-2,0}\,$:
\be
D^{-2,0} \;\Rightarrow \; \nabla^{-2,0} = D^{-2,0} - V^{-2,0}{\cal J}\,.\lb{gCon-2}
\ee
It is related to $V^{2,0}$ by the following appropriate harmonic flatness condition:
\be
\tilde{F} := D^{2,0}V^{-2,0} - D^{-2,0}V^{2,0} = 0\,. \lb{HFd}
\ee
There are also a few additional flatness conditions which follow from the mutual commutativity of the sets $\nabla^{\pm 2, 0}$ and
$\nabla^{0, \pm 2}$, e.g.,
\be
F^{-2, 2} := D^{0,2}V^{-2, 0} - D^{-2, 0}V^{0,2} = 0\,.
\ee
All of them can be shown to follow from the basic harmonic constraints \p{HFa}, \p{HFb} and \p{HFd} (once again, in order to check this,
the general relations \p{Lemma1} and \p{Lemma2} should be used).

Now we are prepared to define a full set of the gauge-covariant spinor derivatives in the considered $\lambda$ frame:
\bea
{\cal D}^{-1,1} &:=& [\nabla^{-2, 0}, D^{1,1}] = D^{-1,1} + D^{1,1}V^{-2,0}\,{\cal J}\,, \lb{Da} \\
{\cal D}^{-1,-1} &:=& [\nabla^{-2, 0}, D^{1,-1}] = D^{-1,-1} + D^{1,-1}V^{-2,0}\,{\cal J} \nn
&=& [\nabla^{0,-2}, {\cal D}^{-1,1}] = D^{-1,-1} + (D^{-1,1}V^{0,-2} +
D^{0,-2}D^{1,1}V^{-2,0})\,{\cal J}\,. \lb{Db}
\eea
The coincidence of the two forms of the same spinor connection in ${\cal D}^{-1,-1}$ can be proved by taking into account
the analyticity of $V^{0,-2}$
and the harmonic flatness condition
$$
D^{-2,0}V^{0,-2} - D^{0,-2}V^{-2,0} = 0\,.
$$

Using the analyticity properties of $V^{2,0}$ and $V^{0,2}$, the harmonic flatness conditions, and \p{Lemma1}, \p{Lemma2},
one can check that the (anti)commutator algebra of these covariantized spinor derivatives coincides with the algebra
of the flat derivatives \p{Dcomm}, up to the
appropriate covariantization of $\partial_t$:
\be
\partial_ t \;\Rightarrow \; {\cal D}_t = \partial_t - \frac{i}{2}D^{1,1}D^{1,-1}V^{-2,0}\,{\cal J}\,.
\ee
By the same token, one can check that ${\cal D}_t$ commutes with all covariantized spinor derivatives,
\be
[D^{1,1},{\cal D}_t] =[D^{1,-1},{\cal D}_t] = [{\cal D}^{-1,1},{\cal D}_t] = [{\cal D}^{-1,-1},{\cal D}_t] = 0\,,
\ee
and with all covariantized harmonic derivatives. It is easy to understand why, in the case under consideration,
there is no analogue
of the gauge-covariant
superfield strengths which are present in similar algebras of the gauge-covariant derivatives in higher dimensions.
The obvious reason is that for the one-dimensional ``gauge field'' ${\cal A}$, ${\cal A}{\,}' = {\cal A} + \dot{\lambda}$, one cannot
construct any analogue of
the covariant field strength. As we have already seen, this ``gauge field'' is the only nonvanishing field in the WZ gauge \p{WZ}.
Substituting the WZ form of $V^{2,0}$ into the flatness condition  \p{HFd}, it is also easy to see that in this gauge
\be
V^{-2, 0}_{WZ} = 2i\theta^{-1,1}\theta^{-1,-1}{\cal A}(t)\,.
\ee

The last topic of this subsection is devoted to the relation with the $\tau$ frame, where the gauge group parameters $\tau$
are harmonic-independent ${\cal N}{=}4$ superfields (in the central basis). The whole set of harmonic flatness
conditions implies that all harmonic connections are expressed through
the bridge $b(z, u, v)\,$,
\be
V^{\pm 2, 0} = D^{\pm 2, 0}b\,, \; V^{0, \pm 2} = D^{0, \pm 2}b\,, \;
b{\,}' = b + \lambda - \tau\,, \; D^{\pm 2, 0}\tau = D^{0,\pm 2}\tau = 0\,. \lb{Brid1}
\ee
The analyticity of $V^{2,0}, V^{0, 2}, V^{0,-2}$ implies that the superfield $b$ is constrained by
\bea
&&(\mbox{a})\;\;D^{1,1}D^{2,0}b = D^{1,1}D^{0,2}b = D^{1,-1}D^{2,0}b = D^{1,-1}D^{0,2}b = 0\,, \nn
&&(\mbox{b})\;D^{1,1}D^{0,-2}b = D^{1,-1}D^{0,-2}b = 0\,, \lb{Brid2}
\eea
where constraints (\ref{Brid2}b) are a consequence of (\ref{Brid2}a). A change to the $\tau$ frame is accomplished by a similarity
transformation,
\be
O_{(\tau)} = e^{-b{\cal J}}O_{(\lambda)} e^{b{\cal J}}\,,\quad \Phi_{(\tau)} =  e^{-b{\cal J}}\Phi_{(\lambda)}\,,\lb{Simil}
\ee
in which $O_{(\lambda)}$ stands for all gauge-covariantized differential operators in the $\lambda$ frame defined above, and
$\Phi_{(\tau)}$, $\Phi_{(\lambda)}$ are superfields transforming by the $\tau$ and $\lambda$ gauge groups, respectively.
Then all harmonic derivatives become ``short'' in the $\tau$ frame:
\be
(\nabla^{\pm 2, 0})_\tau = D^{\pm 2, 0}\,, \;\;(\nabla^{0,\pm 2})_\tau = D^{0, \pm 2}\,, \;\;
\ee
while the gauge-covariant spinor derivatives take the form
\bea
&& ({\cal D}^{1,1})_\tau = D^{1,1} + D^{1,1}b\,{\cal J}\,, \quad ({\cal D}^{1,-1})_\tau = D^{1,-1} + D^{1,-1}b\,{\cal J}\,, \nn
&& ({\cal D}^{-1,1})_\tau = D^{-1,1} + D^{-2,0}D^{1,1}b\,{\cal J}\,, \quad ({\cal D}^{-1,-1})_\tau
= D^{-1,-1} + D^{-2, 0}D^{1,-1}b\,{\cal J}\,, \lb{SpinTau} \\
&& ({\cal D}_t)_\tau  = \partial_t + \left(\partial_t b  -\frac{i}{2}D^{1,1}D^{1,-1}D^{-2,0}b\right){\cal J}\,.\lb{tTau}
\eea
As usual, the form of (anti)commutation relations between gauge-covariant  differential operators is frame independent.

\subsection{The second and third $\lambda$ frames}
As in the previous case, the basic objects of the $\lambda$ frame associated with the {\bf CR} structure \p{CR2}
are harmonic connections
which covariantize the analyticity-preserving harmonic derivatives $D^{2,0}$ and $D^{0,2}\,$:
\bea
&&( D^{2,0}\,, \quad D^{0, 2}) \;\Rightarrow \; \nabla^{2,0} = D^{2,0} - \tilde{V}^{2,0}{\cal J}\,, \quad
\nabla^{0, 2} = D^{0, 2} - \tilde{V}^{0,2}{\cal J}\,, \lb{gCon-22} \\
&& \tilde{V}^{2,0}{\,}' = \tilde{V}^{2,0} + D^{2,0}\tilde{\lambda}\,, \quad
\tilde{V}^{0,2}{\,}' = \tilde{V}^{0,2} + D^{0,2}\tilde{\lambda}\,.
\eea
However, they are subjected to an alternative type of two-theta Grassmann analyticity,
\be
D^{1,1}\tilde{V}^{2,0} = D^{-1,1}\tilde{V}^{2,0} = 0\,, \quad D^{1,1}\tilde{V}^{0, 2} =
D^{-1,1}\tilde{V}^{0, 2} = 0\,.\lb{-V}
\ee
The gauge parameter $\tilde{\lambda}$ also displays this type of analyticity:
\be
D^{1,1}\tilde{\lambda} = D^{-1,1}\tilde{\lambda} = 0 \;\Rightarrow \; \tilde{\lambda}
= \tilde{\lambda}(\zeta_-,u,v)\,.\lb{-V1}
\ee
As in the previous case, connections
$\tilde{V}^{2,0}, \tilde{V}^{0,2}$ obey the harmonic integrability condition
\be
D^{2,0}\tilde{V}^{0,2} - D^{0,2}\tilde{V}^{2,0} = 0\,.
\ee
All other harmonic connections, i.e., $\tilde{V}^{-2,0}$ and $\tilde{V}^{0,-2}$ (where $\tilde{V}^{-2,0}$ is now analytic),
are expressed through these
two basic ones by means of appropriate harmonic flatness conditions. The gauge-covariant spinor derivatives can be constructed  in the
same way as in the previous subsection, starting from derivatives $D^{1,1}, D^{-1,1}$ which are gauge covariant themselves
by virtue of analyticity \p{-V1} of gauge parameter $\tilde{\lambda}$. The full set of harmonic flatness conditions implies a bridge
representation for the harmonic connections
\be
\tilde{V}^{\pm 2,0} = D^{\pm 2, 0}\,\tilde{b}\,, \quad \tilde{V}^{0, \pm 2} = D^{0, \pm 2}\,\tilde{b}\,, \qquad
\tilde{b}{\,}' =  \tilde{b} + \tilde{\lambda} - \tau\,.\lb{-bridge1}
\ee
Here, bridge $\tilde{b}$ is a biharmonic superfield subjected to the constraints
\be
D^{1,1}D^{2,0}\,\tilde{b} = D^{1,1}D^{0, 2}\,\tilde{b} = D^{-1,1}D^{2,0}\,\tilde{b} = D^{-1,1}D^{0, 2}\,\tilde{b}= 0\,, \lb{-bridge2}
\ee
which express another form of analyticity conditions \p{-V}. Note that the harmonic-independent gauge parameter $\tau(t, \theta)$ in
the gauge transformation law of $\tilde{b}$ is the same as in \p{Brid1} because the $\tau$ frame is unique, as distinct from the
three types of $\lambda$ frames related to the existence of three different analytic subspaces in the ${\cal N}{=}4, d{=}1$ bi-HSS.
A change to the $\tau$ frame is accomplished by a rotation similar to \p{Simil}, this time with $\tilde{b}$ instead of $b$.
Once again, the harmonic derivatives become short as a result of this rotation, while gauge-covariant spinor derivatives
take the form
\bea
&& ({\cal D}^{1,1})_\tau = D^{1,1} + D^{1,1}\tilde{b}\,{\cal J}\,, \quad ({\cal D}^{1,-1})_\tau = D^{1,-1}
+ D^{0,-2}D^{1,1}\tilde{b}\,{\cal J}\,, \nn
&& ({\cal D}^{-1,1})_\tau = D^{-1,1} + D^{-1,1}\tilde{b}\,{\cal J}\,, \quad ({\cal D}^{-1,-1})_\tau
= D^{-1,-1} + D^{0, -2}D^{-1,1}\tilde{b}\,{\cal J}\,, \lb{SpinTau2} \\
&& ({\cal D}_t)_\tau  = \partial_t + \left(\partial_t \tilde{b}
-\frac{i}{2}D^{1,1}D^{-1,1}D^{0, -2}\tilde{b}\right){\cal J}\,.\lb{tTau2}
\eea
Since the $\tau$ frame is unique, these expressions for spinor derivatives should coincide with \p{SpinTau},
whence we can conclude that the relations between the bridges $b$ and $\tilde{b}$ are
\bea
&&(\mbox{a})\;\; D^{1,1}b = D^{1,1}\tilde{b}\,, \lb{aRel} \\
&&(\mbox{b})\;\; D^{1,-1}b = D^{0,-2}D^{1,1}\tilde{b}\,, \lb{bRel} \\
&&(\mbox{c})\;\; D^{-2,0}D^{1,1}b  =  D^{-1,1}\tilde{b}\,, \lb{cRel} \\
&&(\mbox{d})\;\; D^{-2,0}D^{1,-1}b  = D^{0,-2}D^{-1,1}\tilde{b}\,. \lb{dRel}
\eea
By applying relations \p{Lemma1}, \p{Lemma2}, one can show that conditions \p{bRel} - \p{dRel} are in fact a consequence of
\p{aRel} combined with the analyticity constraints (\ref{Brid2}a) and \p{-bridge2}.\footnote{It is interesting that
the relations \p{-bridge2} follows from (\ref{Brid2}a), \p{aRel} and \p{cRel}. When restoring $\tilde{b}$ by $b$,
it is more convenient to choose the latter set of constraints as the independent one.} Relation \p{aRel} tells us that
$\tilde{b} -b$ is a three-theta analytic superfield (see Eq. \p{Weak}). This superfield is
still properly constrained by \p{-bridge2}.

In order to make all these statements more explicit, let us restore the bridges $b$
and $\tilde{b}$ for the Wess-Zumino gauge \p{WZ} and show that the latter implies a similar gauge
for the harmonic connections $\tilde{V}^{2,0}$ and $\tilde{V}^{0,2}$.

Using the definition \p{Brid1} it is easy to restore, up to a gauge $\tau$ freedom, the bridge $b$
corresponding to a particular choice \p{WZ},
\be
b_{WZ} =  i\left(\theta^{1,1}\theta^{-1,-1} + \theta^{-1,1}\theta^{1,-1} \right){\cal A}(t_+) -\tau\,,\lb{vWZ}
\ee
where $\tau$ is a superfield which does not depend on harmonics in the central basis, $ \tau = \tau(t, \theta)$,
and we returned to the notation $t_+$ for an argument of $A$ in order to stress that we are still staying in the analytic basis \p{A+}.
Substituting
this expression into \p{aRel} and taking into account \p{-bridge2}, it is straightforward
to restore $\tilde{b}$ up to the appropriate
$\lambda$ gauge freedom,
\be
\tilde{b}_{WZ} = \tilde{\lambda} - \tau  +i\left(\theta^{1,1}\theta^{-1,-1} -
\theta^{-1,1}\theta^{1,-1} \right) {\cal A}(t_+)\,,\lb{vWZ2}
\ee
where $\tilde{\lambda}$ satisfies the alternative analyticity constraints \p{-V1} and is arbitrary otherwise.
In basis \p{A-}, in which the alternative analyticity is manifest, the same object $\tilde{b}_{WZ}$ is written as
\be
\tilde{b}_{WZ} = \tilde{\lambda} - \tau + i\left(\theta^{1,1}\theta^{-1,-1}-\theta^{-1,1}\theta^{1,-1}\right){\cal A}(t_-)
- 2\left(\theta^{1,1}\theta^{-1,-1}\theta^{-1,1}\theta^{1,-1}\right)\dot{{\cal A}}(t_-)\,.\lb{vWZ3}
\ee
Here, the last term is vanishing under the action of $D^{\pm 2,0}$ and $D^{0, \pm 2}$, i.e.,
it belongs to the $\tau$ gauge freedom
(the same additional $\tau$ gauge term also appears in \p{vWZ} after passing to the basis \p{A-}).

For harmonic gauge potentials $\tilde{V}^{2,0}$ and $\tilde{V}^{0,2}$ (in basis \p{A+}) one obtains
\be
\tilde{V}^{\pm 2,0} = D^{\pm 2,0}\,\tilde{b}_{WZ} = D^{\pm 2,0}\,\tilde{\lambda}\,, \quad \tilde{V}^{0, 2} = D^{0,2}\,\tilde{b}_{WZ} =
D^{0,2}\,\tilde{\lambda} + 2i \theta^{1,1}\theta^{-1,1}\,{\cal A}\,. \lb{WZalt}
\ee
Thus, the WZ gauge \p{WZ} for the basic harmonic potentials $V^{2,0}$ and $V^{0,2}$ in the first $\lambda$ frame induces a similar gauge
for the basic harmonic potentials $\tilde{V}^{2,0}$ and $\tilde{V}^{0,2}$ in the second $\lambda$ frame, up to an arbitrary gauge transformation
with the analytic gauge superfunction $\tilde{\lambda}$. This gauge transformation can always be absorbed into an appropriate field redefinition
preserving the alternative analyticity, after which one is left with the pure WZ gauge for $\tilde{V}^{2,0}$ and $\tilde{V}^{0,2}\,$.
This corresponds just to setting $\tilde{\lambda} = 0$ in \p{WZalt}. Alternatively, one could, from the very beginning, work
in the second $\lambda$ frame and
arrive at the WZ gauge for $\tilde{V}^{2,0}$ and $\tilde{V}^{0,2}$ in the same way as expression \p{WZ} was obtained. Then, using
the same identification \p{aRel} - \p{dRel} of the $\tau$ frames, one can show that such a WZ gauge induces expression \p{WZ}
for harmonic potentials in the first $\lambda$ frame, up to an arbitrary gauge transformation with analytic parameter $\lambda(\zeta_+,u,v)$
which can also be absorbed into an appropriate gauge transformation of the involved ``matter'' superfields. To avoid possible confusion,
let us note that it is impossible to  arrange the ``pure'' WZ gauges for the harmonic connections in both $\lambda$ frames {\it simultaneously}:
a set of the WZ connections in one of these frames will always be defined up to an appropriate $\lambda$ gauge transformation.

It remains to consider the $\lambda$ frame associated with the third {\bf CR} structure \p{CR3}.  Once again, the basic quantities are
two harmonic connections, $\widehat{V}{}^{2,0}$, $\widehat{V}{}^{0,2}\,$, which covariantize the analyticity-preserving
harmonic derivatives, namely,
\bea
&& D^{2,0} \;\Rightarrow \; \nabla^{2,0} = D^{2,0} - \widehat{V}{}^{2,0}{\cal J}\,, \quad D^{0, 2} \;\Rightarrow \;
\nabla^{0, 2} = D^{0, 2} - \widehat{V}{}^{0, 2}{\cal J}\,, \lb{gCon-23} \\
&& \widehat{V}{}^{2,0}{\,}' = \widehat{V}{}^{2,0} + D^{2,0}\,\widehat{\lambda}\,, \quad
\widehat{V}{}^{0,2}{\,}' = \widehat{V}{}^{0,2} + D^{0,2}\,\widehat{\lambda}\,.\lb{Tr3}
\eea
These connections, as well as gauge parameter $\widehat{\lambda}$, satisfy the weak analyticity condition
\be
D^{1,1}\widehat{V}{}^{2,0}= D^{1,1}\widehat{V}{}^{0,2} = 0\,, \quad D^{1,1}\widehat{\lambda} = 0\,,\lb{Weak3}
\ee
and are subject to the harmonic integrability condition
\be
D^{2,0}\widehat{V}{}^{0,2} - D^{0,2}\widehat{V}{}^{2,0} = 0\,.\lb{HF3}
\ee
The remaining harmonic connections, $\widehat{V}{}^{-2,0}$ and $\widehat{V}{}^{0,-2}$, are nonanalytic. They are related to
the basic connections $\widehat{V}{}^{0,2}$ and $ \widehat{V}{}^{2,0}$ by appropriate harmonic flatness constraints.
The full set of these constraints implies the existence of a nonanalytic bridge $\widehat{b}$,
such that
\be
\widehat{V}{}^{\pm 2,0} = D^{\pm 2, 0}\,\widehat{b}\,, \quad \widehat{V}{}^{0,\pm 2}=  D^{0, \pm 2 }\,\widehat{b}\,,
\ee
and
\be
\widehat{b}{\,}' = \widehat{b} + \widehat{\lambda} - \tau\,.
\ee
Here, $\tau(t,\theta)$ is the same harmonic-independent $\tau$ frame gauge parameter as in the previous two cases, which expresses
the uniqueness of the $\tau$ frame. The only constraints on $\widehat{b}$ stem from those of analyticity \p{Weak3}:
\be
D^{1,1}D^{2,0}\,\widehat{b} = D^{1,1}D^{0,2}\,\widehat{b} = 0\,. \lb{Weak4}
\ee

A full set of gauge-covariant derivatives in the $\lambda$ frame is constructed, by means of the formulas similar to \p{Da} - \p{Db},
from a single spinor derivative $D^{1,1}\,$. It should not be covariantized in view of the analyticity of the $\lambda$ frame
gauge parameter $\widehat{\lambda}$. As in the previous cases, the $\tau$ frame is achieved by similarity transformation
\p{Simil}, with $\widehat{b}$ instead of $b$. The harmonic derivatives in the $\tau$ frame are short, while the covariant
spinor derivatives and $t$-derivative take the following form:
\bea
&& ({\cal D}^{1,1})_\tau = D^{1,1} + D^{1,1}\,\widehat{b}\,{\cal J}\,, \quad ({\cal D}^{1,-1})_\tau
= D^{1,-1} + D^{0,-2}D^{1,1}\,\widehat{b}\,{\cal J}\,, \nn
&& ({\cal D}^{-1,1})_\tau = D^{-1,1} + D^{-2,0}D^{1,1}\,\widehat{b}\,{\cal J}\,, \quad ({\cal D}^{-1,-1})_\tau
= D^{-1,-1} + D^{0, -2}D^{-2,0}D^{1,1}\,\widehat{b}\,{\cal J}\,, \lb{SpinTau22} \\
&& ({\cal D}_t)_\tau  =
\partial_t + \left[\partial_t \widehat{b}  -\frac{i}{2}D^{1,1}\left(D^{-1,1}D^{0, -2}
+ D^{1,-1}D^{-2, 0}\right)\widehat{b}\right]{\cal J}\,.\lb{tTau22}
\eea

The relations between the bridges $b$ and $\widehat{b}\,$, implied by the uniqueness of the $\tau$ frame, are
\bea
&&(\mbox{a})\;\; D^{1,1}b = D^{1,1}\,\widehat{b}\,, \lb{aRel1} \\
&&(\mbox{b})\;\; D^{1,-1}b = D^{0,-2}D^{1,1}\,\widehat{b}\,, \lb{bRel1} \\
&&(\mbox{c})\;\; D^{-2,0}D^{1,-1}b  = D^{-2,0}D^{0,-2}D^{1,1}\,\widehat{b}\,. \lb{cRel1}
\eea
It is easy to show that relations \p{bRel1} and \p{cRel1}, as well as the weak analyticity constraints \p{Weak4} for $\widehat{b}$,
are a consequence of the constraints \p{Brid2} for $b$ and of relation \p{aRel1}. Substituting the particular expression
\p{vWZ} for $b$ into \p{aRel1}, it is easy to restore the corresponding $\widehat{b}\,$,
\be
\widehat{b}_{WZ} = \widehat{\lambda} - \tau + i\theta^{1,1}\theta^{-1,-1}{\cal A}(t_+)\,,
\ee
where $\widehat{\lambda}(\zeta_3, u, v)$ is an arbitrary three-theta analytic gauge superfunction. Then, for the basic harmonic
connections, we obtain the following expressions:
\bea
&& \widehat{V}{}^{2,0}_{WZ} = D^{2,0}\,\widehat{b}_{WZ} =  D^{2,0}\,\widehat{\lambda} + i\theta^{1,1}\theta^{1,-1}\,{\cal A}(t_+)\,, \nn
&& \widehat{V}{}^{0,2}_{WZ} = D^{0,2}\,\widehat{b}_{WZ} =  D^{0,2}\,\widehat{\lambda} + i\theta^{1,1}\theta^{-1,1}\,{\cal A}(t_+)\,. \lb{WZ3}
\eea

Thus, the WZ gauge for harmonic connections in the first $\lambda$ frame entails a similar gauge for
harmonic connections in the third $\lambda$ frame, up to an appropriate analytic gauge transformation which can be removed by
a redefinition of the involved superfields.
It is worth noting that the WZ gauge for $\widehat{V}{}^{2,0}$, $\widehat{V}{}^{0,2}$ can be independently achieved
just by making use of the $\widehat{\lambda}$ gauge freedom \p{Tr3} and the harmonic constraint \p{HF3}, without any reference to
the first $\lambda$ frame.

The main conclusion of this section is that all three possible $\lambda$ gauge frames are equivalent to each other
under a natural assumption that the $\tau$ gauge frame is unique. Hence, while gauging various isometries
realized on biharmonic superfields, we can choose that $\lambda$ frame which is the most convenient for one or another
purpose. It is worthwhile to note that there is no direct correlation between a choice of the gauge frame and that of the coordinate
basis in the bi-HSS. However, once the $\lambda$ frame has been chosen, it is natural to deal with the basis in which the corresponding
Grassmann analyticity is manifest, i.e. in which the spinor derivatives having no gauge connections are reduced to partial derivatives.
These are the spinor derivatives $D^{1,1}, D^{1,-1}$ in the first $\lambda$ frame, $D^{1,1}, D^{-1,1}$ in the second $\lambda$ frame,
and $D^{1,1}$ in the third $\lambda$ frame. The WZ gauges for the basic harmonic connections covariantizing
the harmonic derivatives $D^{2,0}$ and $D^{0,2}$ have the simplest form just in such coordinate bases, in which these connections
are manifestly analytic.

\subsection{An example of a gauged model in biharmonic superspace}
Diverse models of the ${\cal N}{=}4$ mechanics were obtained in \cite{DI,DI1,DI2} by gauging various symmetries of the one type of the
${\bf (4, 4,0)}$ supermultiplet in the ${\cal N}{=}4, d{=}1$ harmonic superspace with one set of harmonics. Models with nontrivial interaction
can be generated in this way even from a free action of the ${\bf (4, 4,0)}$ supermultiplet. Here, we present an example of gauging symmetries of
the free bi-HSS action \p{qqfree} involving two different types of ${\bf (4, 4,0)}$ multiplets. We shall use the gauge approach
developed in the previous subsections. Since this gauging procedure is manifestly ${\cal N}{=}4$ supersymmetric, the resulting action is
${\cal N}{=}4$ supersymmetric like the initial free action. As distinct from the latter, the gauged action will prove to exhibit a
nontrivial sigma-model-type interaction of component fields.

As was emphasized in \cite{DI}, the symmetries to be gauged should commute with the rigid supersymmetry, otherwise the latter should be
promoted to the local supersymmetry, i.e. to a worldline supergravity. The free action \p{qqfree} enjoys a few symmetries which
meet this requirement.
These are two independent {\it PG}-type SU(2) symmetries realized on the doublet indices $A$ and $\und{A}$, as well as the Abelian shift
symmetries
\be
q^{(1,0)\und{A}}{\,}' = q^{(1,0)\und{A}} + k^{i\und{A}}u^1_i\,, \quad q^{(0,1)A}{\,}' = q^{(0,1)A} + l^{a A}v^1_a\,, \lb{shifts}
\ee
where $k^{i \und{A}}$ and $l^{a A}$ are some constant parameters. The covariance of the defining constraints \p{cq1}, \p{cq2} under these
shifts is evident, and the invariance of \p{qqfree} can be easily proved by representing
$u^1_i =D^{2,0}u^{-1}_i\,, \; v^{1}_a = D^{0,2}v^{-1}_a$,
integrating by parts, and by using the harmonic constraints $D^{2,0}q^{(1,0)\und{A}} = D^{0,2}q^{(0,1)A} = 0$ together with the Grassmann
analyticity conditions in \p{cq1}, \p{cq2}.

All these symmetry groups or some of their subgroups can be gauged using the techniques of the previous subsections. We shall consider the gauging
of some common $\rm{U}(1)$ subgroup of the two {\it PG} SU(2) groups:
\be
\delta q^{(1,0)\und{A}} = \lambda\, C^{\und{A}}_{\;\;\und{B}}\,q^{(1,0) \und{B}}\,, \quad\delta q^{(0,1) A} =
\lambda \,C^{A}_{\;\;B}\,q^{(0, 1) B}\,, \;\; C^{\und{A}}_{\;\;\und{A}} = C^{A}_{\;\;A} =0\,.
\ee
Here, $C^{\und{A}}_{\;\;\und{B}}$ and $C^{A}_{\;\;B}$ are two independent constant SU(2) triplets describing the embedding of the $u(1)$
algebra into a sum of two {\it PG} $su(2)$ algebras. They satisfy the pseudoreality conditions
\be
\overline{( C^{\und{A}}_{\;\;\und{B}})} = -C^{\und{B}}_{\;\;\und{A}}\,, \quad \overline{( C^{A}_{\;\;B})} = -C^{B}_{\;\;A}\,,
\ee
and, without loss of generality, can be chosen in such a way that
\be
C^{\und{A}\und{B}} C_{\und{A}\und{B}} = 2\,, \quad C^{A B} C_{A B} = 2c^2\,, \; c > 0\,,
\ee
where $C^{A B} = C^{BA} = \epsilon^{AD} C^B_{\;\;D}\,$, etc. Using two independent SU(2) rotations, one can choose a frame
in which these  SU(2) breaking tensors take the simple form
\be
C^{\und{1}\und{2}} = i\,, \; C^{12} = i c\,,
\ee
with all other components vanishing.

Now we wish to gauge this $\rm{U}(1)$ symmetry by promoting parameter $\lambda$ to an arbitrary analytic superfunction,
$$
\lambda \; \Rightarrow \; \lambda(\zeta_+, u, v)\,,
$$
and to find a gauge-invariant extension of action \p{qqfree}.

As a first step, we covariantize the defining conditions \p{cq1} and \p{cq2}:
\bea
\mbox{(a)} \;D^{1,1} q^{(1,0)\underline{A}} = D^{1,-1} q^{(1,0)\underline{A}} = 0\,, \quad
\mbox{(b)} \; \nabla^{2,0}q^{(1,0)\underline{A}} =  \nabla^{0, 2,}q^{(1,0)\underline{A}} = 0\,, \lb{cq11}
\eea
and
\bea
\mbox{(a)} \;D^{1,1}q^{(0,1)A} = {\cal D}^{-1,1} q^{(0,1)A} = 0\,, \quad
\mbox{(b)} \; \nabla^{2,0}q^{(0,1)A} =  \nabla^{0, 2,}q^{(0,1)A} = 0\,, \lb{cq22}
\eea
where the gauge-covariant harmonic derivatives and spinor derivatives in the considered (first) $\lambda$ frame are
defined in \p{gCon+2}, \p{gCon-2}, \p{Da}, with
\be
{\cal J}\,q^{(1,0)\underline{A}} = C^{\und{A}}_{\;\;\und{B}}\,q^{(1,0)\underline{B}}\,, \quad {\cal J}\,q^{(0,1)A}
= C^{A}_{\;\;B}\,q^{(0,1)B}\,.
\ee
Second, we replace $q^{(-1,0)\und{A}} = D^{-2,0}q^{(1,0)\und{A}}$ and $q^{(0,-1)A} = D^{0,-2}q^{(0,1)A}$ in \p{Defq-} by
their gauge-covariant analogues $\nabla^{-2,0}q^{(1,0)\und{A}}$ and $\nabla^{0,-2}q^{(0,1)A}\,$.

Then, the gauge-invariant action can be written as
\be
S_{gauge} \propto  \int \mu \left(q^{(1,0)\underline{A}}\nabla^{-2,0}q^{(1,0)}_{\underline{A}}
- q^{(0,1)A}\nabla^{0,-2}q^{(0,1)}_A\right),
\lb{qqgauge}
\ee
with $q^{(1,0)\underline{A}}$ and $q^{(0,1)A}$ defined by the covariantized constraints \p{cq11} and \p{cq22}.

It is now easy to solve Eqs. \p{cq11}, \p{cq22} and to find the component form of the action, using the WZ gauge \p{WZ}:
\bea
&& V^{2,0} = 2i\theta^{1,1}\theta^{1,-1}{\cal A}(t)\,, \; V^{-2,0}
= 2i\theta^{-1,1}\theta^{-1,-1}{\cal A}(t)\,, \; V^{0,2} = V^{0,-2} = 0\,,\nn
&&  D^{1,1}V^{-2,0} = -2i\theta^{-1,1}{\cal A}(t)\,.
\eea
For instance, the gauge-covariant version of solution \p{q1} is obtained via the following substitution in the last component of
$q^{(1,0) \und{A}}$ in \p{q1}:
\be
\dot{f}^{i\und{A}} \; \Rightarrow \; \nabla f^{i\und{A}} =  \dot{f}{}^{i\und{A}} - {\cal A}\,C^{\und{A}}_{\;\;\und{B}}f^{i\und{B}}\,.
\ee
After performing the Berezin and harmonic integrations, one finally obtains the following expression for the component action:
\be
S_{gauge}^{c} \propto \int dt \left( \nabla f^{i\und{A}}\nabla f_{i\und{A}} + \nabla f^{a A}\nabla f_{a A}
+ \frac{i}{2} \psi^{a \und{A}} \nabla \psi_{a \und{A}}+ \frac{i}{2} \omega^{i A} \nabla \psi_{i A} \right),\lb{ActGauge}
\ee
where $\nabla f^{a A} =  \dot{f}{}^{a A} - {\cal A}\,C^{A}_{\;\; B}\,f^{a B}\,,$ etc. Action \p{ActGauge} still respects
the residual $\rm{U}(1)$ gauge freedom under the transformations with parameter $\lambda_0(t) = \lambda |_{\theta =0}$:
\be
\delta f^{a A} =  \lambda_0 \,C^{A}_{\;\; B}\,f^{a B}\,,\; \delta f^{i\und{A}} =  \lambda_0 \,C^{\und{A}}_{\;\;\und{B}}f^{i\und{B}}\,, \;
\delta {\cal A} = \dot \lambda_0 \lb{resid}
\ee
(and analogous transformations for fermionic fields).

One can consider a more general model by adding the Fayet-Iliopoulos (FI) term to the superfield action \p{qqgauge},
\be
S_{gauge} \quad \Rightarrow \quad S_{gauge} -\frac{i}{2}\, \xi\int \mu_{A+}^{(-2,0)}\,V^{2,0}\,.
\ee
The FI term is evidently invariant under the gauge transformation $\delta V^{2,0} = D^{2,0}\lambda\,$. The component action \p{ActGauge}
is modified as
\be
S_{gauge}^{c} \quad \Rightarrow \quad  S_{gauge}^{c} + \xi \int dt \, {\cal A}\,. \lb{ActGauge2}
\ee

By construction, actions \p{ActGauge} and \p{ActGauge2} respect the off-shell ${\cal N}{=}4$ supersymmetry. Taking into account
gauge invariance \p{resid}, which implies that one bosonic field is purely a gauge degree of freedom, the field content
of these actions is still $(8 +8)$.

The gauge field ${\cal A}(t)$ enters \p{ActGauge}, \p{ActGauge2} without derivatives. So it plays the role of the auxiliary field and
can be integrated out by its algebraic equation of motion.
The result is a nontrivial nonlinear $d{=}1$ sigma-model-type action which is still ${\cal N}{=}4$ supersymmetric on shell.
We give here the explicit form of the final nonlinear action in a bosonic limit with all fermionic fields discarded:
\bea
S_{nonl} &\propto & \int dt \left(\dot{f}{}^{i\und{A}}\dot{f}_{i\und{A}} + \dot{f}{}^{a A}\dot{f}_{a A}
- \frac{1}{V}\Pi^2 - \xi^2\,\frac{1}{V} + 2\xi\,\frac{1}{V}\,\Pi \right), \lb{Bosonic}
\eea
where
\be
\Pi = \dot{f}{}^{i\und{A}}C_{\und{A}\und{B}}{f}^{\und{B}}_i + \dot{f}{}^{a A}C_{A B} f^{B}_a\,, \quad V = {f}^{i\und{A}}{f}_{i\und{A}} +
c^2{f}^{a A}{f}_{aA}\,.
\ee
The bosonic action contains a nonlinear sigma-model part (the terms bilinear in time derivatives), a potential term ($\sim \xi^2$)
and the WZ-type coupling to an external gauge potential (the term $\sim \xi$).

To summarize, we have started from action \p{qqfree} which is a sum of free actions of two ${\bf (4, 4,0)}$ supermultiplets
and, after gauging its $\rm{U}(1)$ isometry, arrived at the action
with a nontrivial self-interaction mixing the fields from both multiplets. The final action still respects
the local $\rm{U}(1)$ symmetry \p{resid},
and one can fully fix it by making one of the bosonic target coordinates zero. So, finally, we have a nonlinear sigma-model-type action
with a seven-dimensional bosonic target manifold, involving nontrivial WZ  and potential terms. Let us note that the original
${\cal N}{=}8$ supersymmetry
of action \p{qqfree} is not preserved by the gauging procedure; the gauged actions possess only ${\cal N}{=}4$ supersymmetry. We note also that
the target metric and potential in \p{Bosonic} are singular at the point $f^{i\und{A}} = f^{a A} =0\,$, and thus one is led to assume that
these bosonic fields have some nonzero background values.\footnote{Actually, this singularity can be removed if the original $\rm{U}(1)$
symmetry is chosen to
have an admixture of some one-parameter transformation from the shift symmetries \p{shifts}.} The more detailed study of geometric
properties of this model, as well as of some other ones associated with different gaugings of symmetries realized on
the superfields $q^{(1,0)\und{A}}$ and $q^{(0,1)A}\,$, will be performed elsewhere. The options of special interest
are those in which one of the two SU(2) $PG$ groups, or their diagonal
SU(2) subgroup, are gauged. The resulting ${\cal N}{=}4$ supersymmetric models with five-dimensional target bosonic manifolds could reveal
a close relationship to supersymmetric mechanics with the Yang monopole as a target (see the recent paper \cite{TopNer},
the authors of which also proceed from an ${\bf (8, 8, 0)} = {\bf (4, 4, 0)}\oplus {\bf (4, 4, 0)}$ multiplet
of the ${\cal N}{=}8$ supersymmetry).

\setcounter{equation}{0}
\section{Conclusions and outlook}
In this paper, we worked out the basic elements of a new
systematic superfield approach to models of the ${\cal N}{=}4$
supersymmetric mechanics based on the concept of biharmonic
superspace. In this approach, both SU(2) $R$-symmetries
of the ${\cal N}{=}4, d{=}1$ super Poincar\'e algebra are ``harmonized'' and prove
to be on equal footing, thus allowing a joint description of
the off-shell ${\cal N}{=}4$ supermultiplets which are mirror to
each other. Here, we limited ourselves to presenting a few particular
examples of how useful this approach is for the ${\cal N}{=}4$
mechanics model-building, leaving  a more extensive study of its
possible applications in this area for the future. It should be
mentioned that, until now, only the ${\cal N}{=}4$ mechanics models
based on one type of ${\cal N}{=}4$ supermultiplets were mostly
studied; having the bi-HSS approach, one will be able to explore the
more general models combining these supermultiplets together with
their mirror counterparts. In the case with one sort of multiplets,
the whole variety of ${\cal N}{=}4$ mechanics models can be
generated by appropriate reductions (related to the notion of
the ``automorphic duality'' \cite{GR0}-\cite{T}) from models
associated with the root multiplet ${\bf (4,4,0)}$ \cite{root}.
In the superfield language, these reductions amount to diverse
gaugings in the standard harmonic ${\cal N}{=}4, d{=}1$ superspace
\cite{DI,DI1,DI2}. In the case where both sorts of ${\cal N}{=}4$
multiplets are incorporated, we expect a similar phenomenon with a
pair of the complementary biharmonic supermultiplets $q^{(1,0)
\und{A}}, q^{(0,1)A}$ as the root ones and with the gauging procedure
presented in Sect. 5 as a generalization of that of Refs.
\cite{DI,DI1,DI2}. As shown in a recent paper \cite{Calo}, the
superfield gauging approach is efficient for deriving novel ${\cal
N}=4$ superextensions of some integrable Calogero-type $d{=}1$ models.
The inclusion of pairs of mutually mirror ${\cal N}{=}4$
multiplets into the scheme of \cite{Calo} within the bi-HSS approach
could result in an essential enlargement of this class of
supersymmetric $d{=}1$ systems.

The bi-HSS approach can also, presumably, provide new opportunities in a different circle of problems, for instance those related
to supersymmetric integrable systems of the Korteweg-de Vries (KdV) type. It was found in \cite{DI0} that the harmonic ${\cal N}{=}4, d{=}1$
analyticity underlies an ${\cal N}{=}4$ super KdV equation. The second Hamiltonian structure
of the latter, namely, the small ${\cal N}{=}4$ SU(2) superconformal algebra, in the $d{=}1$ harmonic superspace approach
with one set of the SU(2) harmonics $u^\pm_i$ is naturally represented by an analytic  supercurrent
$J^{++}(\zeta, u^\pm)$ satisfying the harmonic constraint $D^{++}J^{++} = 0$. It collects the Virasoro, superconformal,
and SU(2) affine currents as independent components in its $\theta$ expansion. The direct bi-HSS analogue of this superfield
is $J^{(2,0)}(\zeta_+,u,v), \;D^{2,0}J^{(2,0)} = D^{0,2}J^{(2,0)} = 0$. A new feature of the bi-HSS is the possibility
to define a mirror supercurrent $J^{(0,2)}(\zeta_-,u,v), \;D^{2,0}J^{(0,2)} = D^{0,2}J^{(0,2)} = 0\,$. It can
generate another ${\cal N}{=}4$ SU(2) superconformal algebra which, together with the first one, would yield
the large ${\cal N}{=}4$ ${\rm SO}(4)\times {\rm U}(1)$ superconformal algebra as a closure.
These two supercurrents can be used as the basic entities
of a new ${\cal N}{=}4$ super KdV system with the large superconformal algebra as the second Hamiltonian structure.
Such a superfield extension of the KdV equation has not been constructed yet. On top of that, one more notable
feature of the ${\cal N}{=}4$ bi-HSS,
viz. the existence of the three-theta analytic subspace \p{Asubs3}, can hopefully be utilized as a new tool in attempts
to construct ${\cal N}{=}4$ superextensions of Zamolodchikov's $W(3)$ algebra. This problem remains open, too.

\section*{Acknowledgements}
E.I.  thanks the Institute of Physics in Prague for the kind hospitality extended
to him during the course of this study. His multiple visits to Prague were supported by grants from the
Votruba-Blokhintsev Program and by the TEORIE grant.
He also thanks the Leibniz University of Hannover for warm hospitality at the final stage
of this study. His work was supported in part by RFBR Grants No. 09-02-01209 and No. 08-02-90490, DFG
Project No. 436 RUS/113/669, INTAS Grant 05-1000008-7928, and a grant from the Heisenberg-Landau Program.

\end{document}